# POSSIBILITY OF THE EFFECT OF ANISOTROPIC TRANSPARENCY IN SEMI-CONDUCTOR STRUCTURES.

## Lazarev S.G.
Sarov, Russia.
s.g.lazarev@gmail.com

*The present paper examines the possibility of asymmetric effect of the adjacent to barrier physically different media, or anisotropic structure of the barrier itself, on probability of the oppositely directed transitions of current carriers through the potential barrier. It is shown, that in the above cases transparency of the barrier can be anisotropic. Consequently, there is a possibility that the chaotic thermal motion of particles may self-transform into the ordered motion. Stationary state of such systems turns to be different from thermodynamical equilibrium, which, in its turn, results in possibility of the effective work due to spontaneous cooling of the barrier region. Thermodynamics of these processes is being considered herein.*

## Introduction.

Several methods, using which thermal energy can be transformed into effective work are known. Transformation goes, for instance, due to Zeebeck effect /1, page 321/, underlying the work of thermocouples. In this case, to have electromotive force (EMF), the pair of two different conductive materials are to be maintained in contact at varying temperatures. EMF exhibits itself also when temperature gradient is maintained in a homogeneous conductive material /2, page 18/. EMF is stipulated here both by the diffusion of the carriers, originating from the hotter region, and the flux of phonons, transporting the same. It is essential, that transformation of thermal energy into the effective work takes place in this and another known cases only in thermodynamically non-equilibrium systems and is accompanied by entropy gain. Conversion of thermal energy into the effective work in equilibrium systems (i.e. realization of the so-called "perpetuum mobile of the second kind") is deemed to be impossible. This postulate is reflected in the second law of thermodynamics.

Phraseology of the second law has several options. For instance, in /4, page 60/ analysis of 18 fundamental statements is presented. One of those is Kelvin's postulate, according to which it is impossible to make such a machine, operating in a cyclic mode, the only effect of which could be extraction of heat from a reservoir and commitment of the equal to the same quantity of work /5, page 109/. The modern wording is as follows: "if in a certain moment of time entropy of the closed system is different from the maximum, then in the follow-up moment entropy does not subside: it either grows or stays constant" /6, page 45/. The first wording reflects the practical aspect of implementation of thermal energy, whereas the second dwells on the statistic character of description of macroscopic systems.

*Utopian "perpetuum mobile of the second kind".*

Numerous attempts to develop the new methods of transformation of equilibrium thermal energy into effective work were undertaken and published in scientific and patent literature. In one of those papers /7/, for instance, the author speaks about embodying the cyclic process, implicating the stages of stretching, cooling, compression and heating of a spring, so, that due to the growth of its rigidity during cooling, the effective work is done. In /8-10/ he discusses straightening of fluctuations in the structures featuring non-linear volt-ampere characteristic.

It can be pointed, with respect to /7/, that in the approximation, according to which rigidity of the spring depends on temperature, its adiabatic deformation would not be isothermal, as opposed to what it said in /7/. This can be easily get, if dependence of the modules of overall compression and shear, as well as of thermal expansion factor on temperature are considered, as they are expressed in the form of free energy of a body /14, page 28/. In this case the difference of temperatures of the heater and the cooler, as well as the amplitude of deformation happen to be bound by inequality, which provides for lower efficiency of such a process, than that of Carno cycle.

In /50/ it is proposed to transform heat into effective work using some thermally inhomogeneous systems, and it is stated, that having the self-organization phenomenon at hand, one can obtain the negative production of entropy. Criticism of what is postulated in this paper is presented in /51/.

In /8-10/ several schemes of embodying the patented method are presented. The peculiarity, similar to all of them is, that using the non-linear elements, for instance diodes, a physically microscopic volume is isolated. Due to thermal fluctuations, accompanying exchange by the carriers of the current inside the isolated volume, with the external medium via non-linear elements, the electric charge and, consequently, the potential of the isolated volume become alternating. It seems here, ex facte, that detection of fluctuations will take place here, resulting, naturally, in the constant component of current to occur. However, as the analysis shows, the increasing intensity of transitions of current carriers through the non-linear element in the forward for the same direction from the charged isolated volume is compensated by the decreased intensity of transitions in this direction through another non-linear element, provided that the isolated volume is non-charged. As a matter of fact, in the first case the charged system commits its work with the electron being considered, whereas in the second case the electron commits the equal work aimed at charging the system.

The above factors considered for in the expression for probability of thermal fluctuations, the average current and the energy generated turn to be equal to zero. Absence of current generation in equilibrium conditions also is recorded for the devices, patented in /11-13/, as well as in another well-known to those skilled in the art technical solutions.

In general it is worth saying, that the methods of implementation of equilibrium thermal energy for the effective work to be performed are either unknown or "restricted" by the second law of thermodynamics.

*Pseudo-perpetuum mobiles.*

Beside the utopian "perpetuum mobiles", in which the processes, violating the first and the second laws of thermodynamics are implemented in the area, where they can be correctly used (i.e. "perpetuum mobiles" of the first and the second kind), the so-called "pseudo-perpetuum mobiles" exist. One of the widely met everywhere mechanisms of this type is the device for winding the mechanic watches, converting the essentially chaotic movements of human hand into strictly directed movement of pointer. Also known are the devices, based on the principle of fluctuation of atmospheric pressure and temperature, etc /15, page 221/, providing the directed motion of the working body. It is evident, that these devices do not need any additional energy sources to operate them and can work, virtually, without any limits. The peculiarity of these pseudo-perpetuum mobiles is implementation of mechanical "anisotropic barriers", like ratchet gear, providing asymmetry of the working body with respect to the exterior effects.

Miniaturization erases to some extent the border between macroscopic (for the present scales of the device) and microscopic (thermal) fluctuations of physical parameters. The problem of the minimum size of the device, converting the chaotic fluctuating effect of the environment into the directed motion of the working body is still acute /48, 49/.

The example of the so-called "perpetuum mobile of the second kind" by Thomson-Planck is known in statistical physics, - the device, the effect of which exhibits as periodical positive work due to cooling of only one body, without any changes caused in another bodies /16, page 53/. Such a process is only possible in the systems of nuclear and electronic spins. Their thermodynamic "extraordinary" feature is stipulated by the limited on the upper side energy spectrum, which makes it possible to obtain negative temperatures.

## Spatial asymmetry of micro-transitions.

The ability of symmetric influence of the adjacent to barrier physically different media, or the anisotropic structure of the barrier itself, on probability of opposite transitions of current carriers through the potential barrier is worth examining. Anisotropic transparency of the barrier to current carriers results in the ability of transformation of energy of the chaotic thermal motion of electrically charged particles into the energy of their directed motion. Such a transformation, in its turn, gives the chance for the effective work to be made due to cooling of the barrier region. The present thermodynamical paradox must be analyzed in every detail.

*Currently known factors, limiting adaptability of the second law of thermodynamics.*

Spontaneous cooling of the barrier region, i.e. exhibition of its anisotropic transparency, seemingly contradicts the second law of thermodynamics. However, any physical law has the restricted area of application. Applicability of the second law of thermodynamics is also known to be restricted with respect to big gravitating systems, on the hand, and to the motion of singular bodies, on the other hand /6, c. 46, 16, c. 19/.

As it is believed, the processes, considered in the present paper, due to the reasons mentioned below, also fail to be meet the conditions of the correct application of the above physical law.

It has to be mentioned, that despite the numerous theoretical studies and quite credible experimental proofs, the problem of physical substantiation of the law, governing monotonous growth of entropy, still remains under-examined /6, page 45/. Until present the second law of thermodynamics has not been substantiated in terms of classical quantum mechanics or the field theory /6, page 44-49/. That is why, strictly speaking, it is still being viewed only as some trustworthy scientific postulate /5, page 109/.

The second law of thermodynamics was formulated 150 years ago, when the averaged properties of gigantic assembles of particles exhibited themselves practically in every processes observable. The only microscopic effect, in which averaging was not quite complete, was the Brownian motion (1827). To date, as microelectronic technologies develop, the dimensions of certain devices tend to the atomic scale. As a result, the structures, in which interaction of separate particles and the complete ensemble of them is combined in a more sophisticated manner, than in Carno machine, became practical.

Theoretical studies of transition of current carriers in semi-conductor anisotropic structures revealed, that some factors exist, stipulating asymmetry of the intensities of the direct and backward transitions of particle systems, when they move thermally between the groups of micro-states, corresponding to the state, when the particles are on the different sides of the potential barriers. In particular, the time of relaxation of medium is one of such factors /17/. Asymmetry of micro-transitions poses certain limitations on the level of anisotropy in stationary state and, correspondingly, on applicability of the second law of thermodynamics, as it describes these systems.

With asymmetry of micro-transitions considered for, the stationary state of the above systems becomes thermodynamically "extraordinary" and stabilizes at incomplete thermodynamic equilibrium, acknowledged as the equality of thermodynamical parameters of all its sub-systems. Stationary values for the levels of Fermi sub-systems become different as well.

### The effect of the adjacent media on transparency of the barrier.

Potential barriers of various types for current carriers in solid bodies are widely used in electronics. These are, for instance, semiconductor heterojunction structure, Schottky barriers, metal-dielectric- semiconductor systems, etc. The properties of these barriers were studied in numerous theoretical and experimental endeavors /18, 19/.

The significant aspect of tunneling theory is the effect, the medium, surrounding the barriers, poses on its transparency. It is known, for instance, that interaction of electron, crossing the potential metal - semiconductor barrier, with the induced electric charge in the metal, results in the reduction of height of the potential barrier /2, page 240/. This phenomenon is known as Schottky effect. For immobile charged particles, located close to the conductive surface, this effect was studied quite exhaustively. The power, effecting the electric charge, exhibits also near the interface of the two non-conductive media with different dielectric permittivity /20, page 61/. Thus, the formalism of dielectric permittivity with consideration for spatial dispersion and using the Green function, is applicable in theories of absorption, catalysis, etc, as well as when studying inter-electronic actions in layered high-temperature superconductors /21/.

In /22, page 264/ being studied is the question: what, in fact, has to be understood as the value of dielectric permittivity $\varepsilon$ when considering the process of electronic exchange through the barrier. The value $\varepsilon$ may differ from the statistic one, as it is pointed, since during emission the time of flight of electron through the barrier region may be less than that, required for polarization of semiconductor. However, it is shown further, that dielectric permittivity of silicon practically equals to the statistical value. The appropriate theoretical assessments and experimental data are presented.

In some papers the dynamic effects, accompanying the tunneling process are considered, resulting in dependence of the height of the potential barrier on the energy of the tunneling particle /23-25/. In /23/, for instance, the expression for the potential, in which the point particle, crossing the gap between two (different in general case) conductive media was drawn, using the formalism of dielectric permittivity (the processes of dissipation not considered for). In /24, 25/ the similar calculations were made by path integrals methods, with consideration for interaction of the particle with surface plasmons in the media, separated by the barrier. The expressions, demonstrating dependence of the barrier height on the particle energy were drawn, which provides the refined expression for probability of tunneling.

### The effect of relaxation on barrier height.

The effect of initial conditions on probability of transitions through the barrier in anisotropic structures with finite time of relaxation was not studied exhaustively until now. In the meantime it has to be born in mind, that the charge carriers in plasma, in particular in solid state plasma, are represented by quasi-particles, coated by correlation charge, stipulated by both Coulomb and exchange interaction. It is significant, that the parameters of the coat, such as screening radius, for instance, or correlation energy, are different for the different media, as the characteristics of quasi-particle, consequently. It is not less significant, that both formation of correlation charges, when the charged particles enters the medium, and their vanishing, when the particles escapes, are not the instant processes, - they occur during the time of Maxwell relaxation or the period of the natural oscillations of plasma (the lattice relaxes at the time approximately equal to the period of atomic oscillations). Finally, the time of transition of the particle through the barrier (tunneling time), which still is under discussion and had not been identified strictly until now /26/, nevertheless depends on relation of its energy and the parameters of the barrier and, can be bigger or less than the characteristic times of the corresponding degrees of freedom of media, with which the tunneling particle interacts.

The above factors prove, that when interaction of the charged particles, crossing the barrier with correlation charges is considered in case of physically different media, separated by the potential barrier, its local transparency (barrier height) turns to be dependent on pre-history of the process (initial conditions of transition through the barrier) and, correspondingly, on the direction of motion of the charged particle.

Thus, if certain parameters of media can be selected, the conditions can be achieved, when the average height of barriers for the opposite directions of carrier motion may become significantly different.

It is known, that with the generally        assumed equality of barrier height for the oppositely directed particles, stationary states of the separated by the barrier sub-systems, characterized by equality of exchange currents, settle with the equal thermodynamical parameters. Consequently, it can be expected, that in case, when the average height of the barrier for the opposite directions of carrier motions are different, stationary states of sub-systems will settle with the shift of thermodynamical parameters, in particular, with the shift of the levels of electro-chemical potential, which is proportional to the difference of the average heights of the barrier.

When the media are locked by an exterior electric circuit via the identical barrier, the stationary state will settle in the resulting system with the shifted thermodynamical parameters of sub-systems, although with a zero complete current over the present closed electric circuit. However, when media are closed by means of an exterior circuit without barriers, or featuring more transparent barriers, stationary state in the system will be achieved only at the incomplete non-zero current, due to different stationary shift of Fermi levels at each of the contacts. Thus, the electric current is energized in the circuit, which may include the effective load. As a consequence, conversion of energy of chaotic (thermal) motion of current carriers into the energy of their directed flux can be achieved, which may be used for commitment of effective work, as well as for cooling of the adjacent to the transformer medium.

It is worth mentioning, that due to anisotropy of the barrier height, the energy spectrum of counter-propagating fluxes of particles has to be a little different. As a consequence, even at the completely zero current energy transfer through the barrier will occur, compensated by heat transfer over the ion lattice. Peltier effect is the closest in its essence to the present process /20, page 147/. When considering the energy flux in the circuit, it can be revealed that the region in the vicinity of anisotropic barrier will be cooled, and the corresponding energy release will take place on the resistor of the exterior load. The total energy release of the whole system is equal to zero, according to the law of energy conservation.

When the barrier temperature or the average temperature of the contacting media tends to the absolute zero, Fermi levels tend to the unified for all the system Fermi energy, independent from the parameters of the barrier, since electrons can transit only into the unoccupied states. Current generation in this case stops.

Asymmetry of probability of the opposite transitions through the barrier in anisotropic semiconductor structures can be shown both in the classical approximation and the quantum description of this phenomenon.

**The effect of relaxation on the height of Schottky barrier.**

Let us assume, that the barrier (fig.1) is insufficiently wide (Schottky barrier with impurity-doped semiconductor), such, that electron can be viewed as the classical point particle /2, page 240/ and the effect can be considered within the ranges of the accepted in Schottky theory formalism. The Schottky theory can be refined, as interaction of the electron, moving through the barrier, is considered not only for the conductive metal surface, but the conductive medium of semiconductor. In case, when interaction of electron with the conductive medium of semiconductor can be described similarly to the interaction with metal in approximation of electrostatics, for instance by the method of image charges, solution of the problem of the     second conductive surface will not cause any trouble /18, page 131/. However, usually such an interaction is not considered due to comparative negligibility of the effect.

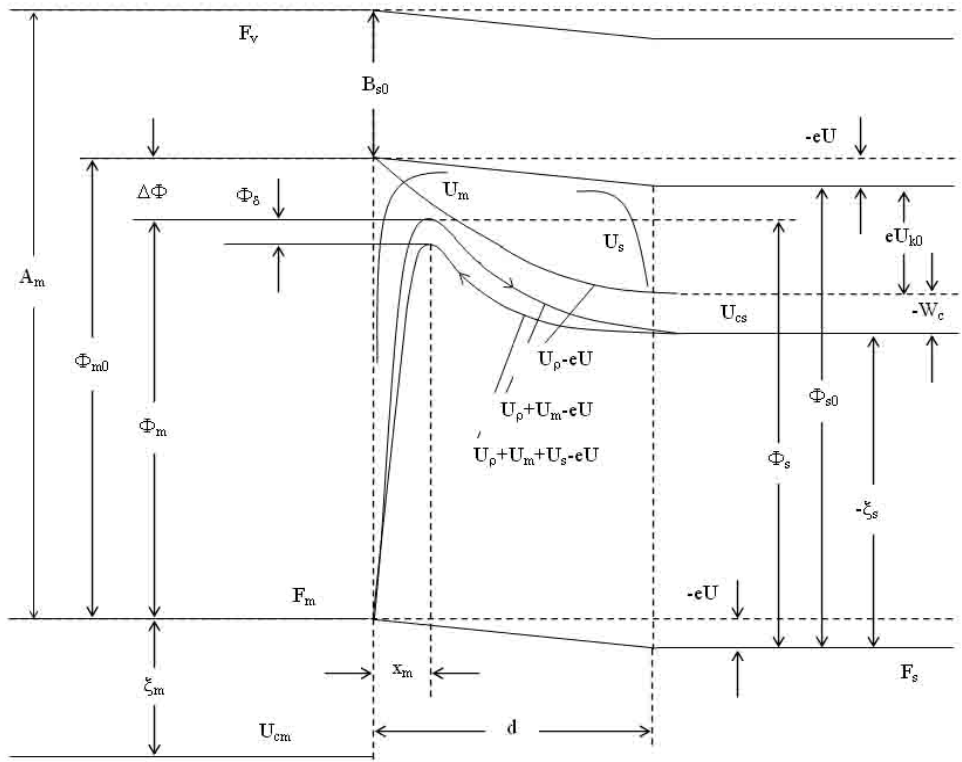

Fig.1

Fig. 1. Energy diagram of the blocking junction of the two media (Schottky barrier). Index m stands for the parameters of the rapidly relaxing medium (metal), index s – for the slowly relaxing medium (semiconductor). $F_m$ and $F_s$ – are Fermi levels, $U_{cm}$ and $U_{cs}$ – energies of the bottom of conductive band, $F_v$ – the rest energy of electron in vacuum, $\xi_m$ and $\xi_s$ – chemical potentials, $U_m$ and $U_s$ – energy of image force of media, $W_c$ – correlation energy, $A_m$ – work of escape into vacuum from fast-relaxing medium, $B_{so}$ – electronic similarity of slowly-relaxing medium (work of escape into vacuum from the bottom of conductive band without correlation energy considered), $\Phi_{mo}$ and $\Phi_{so}$ – work function without image force considered, $\Phi_m$ and $\Phi_s$ – work function with image force considered, $\Phi_\delta$ - difference of barrier heights with energy considered and not considered for, $U_s$, $U_\rho$ - profile of energy barrier without mirror image force considered, U – electric potential of slowly-relaxing medium relative to the fast-relaxing one, $x_m$ – coordinate of the maximum potential barrier relative to the medium potential barrier, d – barrier width (region, depleted with current carriers), $eu_{ko}$ – barrier height relative to the bottom of the conductive band of slowly-relaxing medium at zero difference of electric potentials between media without consideration for image force and correlation energy, $eu_k$ – barrier height relative to the bottom of conductive band of slowly-relaxing medium at zero difference of electric potentials between the media. The arrows point to the directions of transitions.

To correctly use approximation of electrostatics, the time of Maxwell relaxation, in particular, has to be far less, that the time needed for electron to overcome the barrier. In case of metals this condition is fulfilled     well, as a rule. As for semiconductors, their electric conductivity may vary in a dramatically wide range. It may be selected such, that the time of relaxation of semiconductor medium will be comparable with characteristic time of electronic passage.

Let the current carrier move through the barrier made of metal towards semiconductor, featuring quite a big time of relaxation, such, that at the period, starting from takeoff of the carrier towards the barrier region up to the moment it reaches its peak, the "mirror" charge in semiconductor would not have enough time to form. The height of the barrier in this case will be defined by electrostatic interaction of the current carrier only with the free charges of the metal.

Let us consider the opposite case, when the current carrier moves from semiconductor towards metal. In this case semiconductor medium is assumed to be originally relaxed, - the carrier is surrounded by the coat represented by correlation charge. The height of the barrier in this case is defined by interaction with electric charges in both media, or the medium with higher summarized charged, to be more precise, and, correspondingly, turns to be a little lower, than in the first case.

It is worth mentioning, that formally in a stationary "smooth" potential, which in the proximity of the peak may be considered parabolic, the particle with full energy, which is strictly equal to the height of the potential barrier, reaches the same within the infinite time, which will seemingly result in equal relaxation of media for the oppositely directed particles with the limiting energy, thus removing anisotropy.

However, the charged particle, as it moves in the barrier region in the proximity of the conductive surface, excites the alternating current in the conductive medium, thus dissipating some part of its energy. This results in friction force, depending on speed to occur. Description of such an effect is presented in /20. page 538/. That is why for the particle to overcome the barrier some more energy will be needed (as much as the dissipation losses, depending on the properties of the media), i.e. more than the potential barrier height. Consequently, the effective height of the barrier for the opposite directions of particle movement will be different, as relates to different adjacent media.

Above all, it is quite evident, that with thermal fluctuations of potential taken into consideration, the time period, during which the electron, featuring any initial pulse, stays near the peak of the barrier becomes finite.

Thus, the energy potential occurs, in which the current carriers can move only in one direction. Compensation of the excessive current of the carriers throughout the above energy interval takes place at the corresponding shift of Fermi levels.

### *A transparent in terms of tunneling barrier.*

Let us now consider another extreme case of the narrow, if compared with electron wavelength barrier (tunnel barrier), which may be represented here by semiconductor heterojunction. In this case anisotropy of the barrier height also is present, stipulated by interaction of the tunneling particle with the relaxing correlation charges, which, at the closer look, should be considered as the combination of quantum fields of quasi-particle excitations. Above all, some of the factors, described below, make a certain contribution to anisotropy of the tunneling process.

To describe anisotropy of the barrier, interaction of the tunneling electron with quasi-particle excitations in the adjacent to the barrier media should be taken into consideration obligatory. Should this interaction be neglected, the wave function of electron becomes coherent for all the volume under consideration. In this situation Kramers theorem becomes true /27, page 102/, according to which asymmetric with respect to operation of spatial reflection band of energy are impossible. The theorem is proved by complex conjugation of Schredinger equation. Hamiltonian of the equation, still remaining real, undergoes auto-transformation, although the wave vector of electron changes the sign to the opposite one, which is true for any symmetry of crystal. Since transition to the complexly conjugated Schredinger equation is equivalent to change of the sign of time, here we use, in fact, the symmetry of Schredinger equation relative to inversion of time.

Consequently, to have spatial asymmetry, the situation, in which Hamiltonian of a separate electron is asymmetric with respect to inversion of time have to occur. Such a situation occurs, when transition of electron into the alternative state is significantly effected by interaction of the same with the system, described rather by statistical equations, non-symmetrical to time inversion (interaction with statistical ensembles of particles), than the mechanical ones (quantum-mechanical, - interaction with non-separate particles).

Above all, irreversibility of electron tunneling through the barrier is intrinsic to the quantum character of this phenomenon. Essentially, in this case substitution of the wave function, corresponding to presence of electron on one side of the potential barrier, for the wave function, corresponding to presence of electron on the other side of the barrier, takes place, instead of the change of wave function, described by Schredinger equation. Thus, in this instance, "reduction of the wave package" occurs /41, page 64/: after tunneling the state of electron changes irreversibly. Discussion of the acute problem of "reduction of the wave package" is presented, for instance, in /53/.

In certain sense the adjacent to the barrier media play the role of measuring devices. With interaction absent, the states of electron are possible, possessing equal probability of being present (up to the moment of measurement) at the different sides of the barrier. However, interaction of electron with a measuring tool (a special probe or a crystal lattice – it does not matter) on one side of the barrier make the probability of identification of electron on the other side of the barrier, when measurement is repeated, very low (at typical conditions). Irreversibility of this type inserts physical inequivalence of both time directions into the tunneling process /42, page 41/.

The effect of occurrence of stationary, thermodynamically non-equilibrium state in hetero-structures is stipulated by inelastic processes, accompanying change of energy of the tunneling particle.

Certain effects of inelastic electron dissipation during tunneling in the barrier are known /see, for instance 29/. Generally speaking, manifestation of the tunneling effect, accompanied by various excitations, was observed, for instance, in volt-ampere characteristics of p-n transitions /19. page 262/.

To qualitatively analyze interaction of a particle with the "takeoff" medium, one may consider the energy of a quasi-particle, in the classical (simplified) approximation, as the sum of proper energy of electron, energy of the polarized region (correlation coat) and energy of electron interaction with the "correlation coat" /45, page 38/. When going through the barrier, electron is "torn off"      from its correlation coat. Energy of their interaction decreases in this case. When the particle leaves the polarized region infinitely long, both energy of their interaction and the energy of the polarized region gradually decrease down to zero. On the other side, when the particle infinitely quickly leaves for infinity, interaction energy vanishes, but the energy of the polarized region stays as it is. Further on this energy dissipates into heat. The share of the dissipating into heat energy of the tunneling electron have to be dependent on correlation between tunneling time and the characteristic times of formation and dissipation of the correlation coat, as well as the correlation between correlation radius and barrier thickness. Physical processes here are similar to the ones, happening during photo-excitation of polaron /37, page 86; 45, page 61/. However, spatial anisotropy makes inelastic tunneling significantly different from photo-excitation of polaron.

Figs. 2 - 6 are diagrams, illustrating the tunneling process with the effective masses of electron on different sides of the barrier taken into consideration, stipulated by peculiarities of the crystal lattice, the process of electron-photon interaction also taken into consideration. Also presented are the corresponding to the diagrams spectral densities of current for temperatures T = 100K, 200K, 300K. Above all, the functions of distribution are presented for electrons, corresponding to these temperatures. Similar to Feinmann diagrams, the pointer corresponds to Green function of the operator of fermion (electron) field. The difference between effective masses of electrons is illustrated by alternating in thickness pointers. The dotted line corresponds to Green function of bosonic (phonon) field. Difference of dispersion characteristics of bosons is also illustrated by alternating line thickness. The tip of the diagram corresponds to the tunnel transition and is depicted as a circle. Thus, the enlisted diagrams graphically represent the following processes:

Diagrams a) and b) fig. 2 symbolically depict the direct and the reverse processes of elastic tunneling through the barrier. Fig. 2) – tunneling of the free electron through the barrier from the left to the right. The value of the matrix element of this process is precisely equal to the reverse one, depicted in the diagram in fig. 2(b)

Figs. 3(a), 4 (a), 5 (a) and 6 (a) – tunneling of electron from the left to the right together with the processes of radiation and absorption of phonons of the first order.

Figs. 3 (b), 4 (b), 5 (b) and 6 (b) – tunneling of electron from the right to the left together with the processes of radiation and absorption of phonons of the first order. $j_{rl}^{0}, j_{lr}^{0}, j_{rl}^{l+}, j_{rl}^{l-}, j_{rl}^{r+}, j_{rl}^{r-}, j_{lr}^{l+}, j_{lr}^{l-}, j_{lr}^{r-}, j_{lr}^{r+}$ - densities of the tunneling current, corresponding to the abovementioned processes.

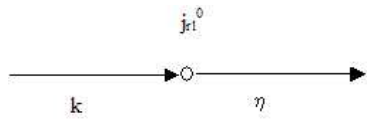

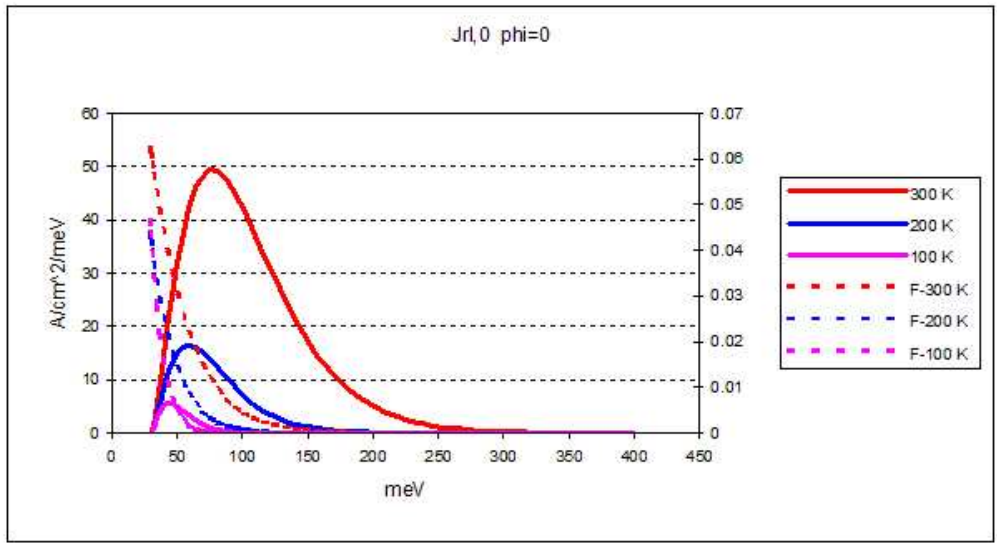

a)

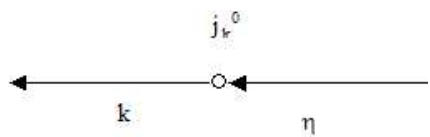

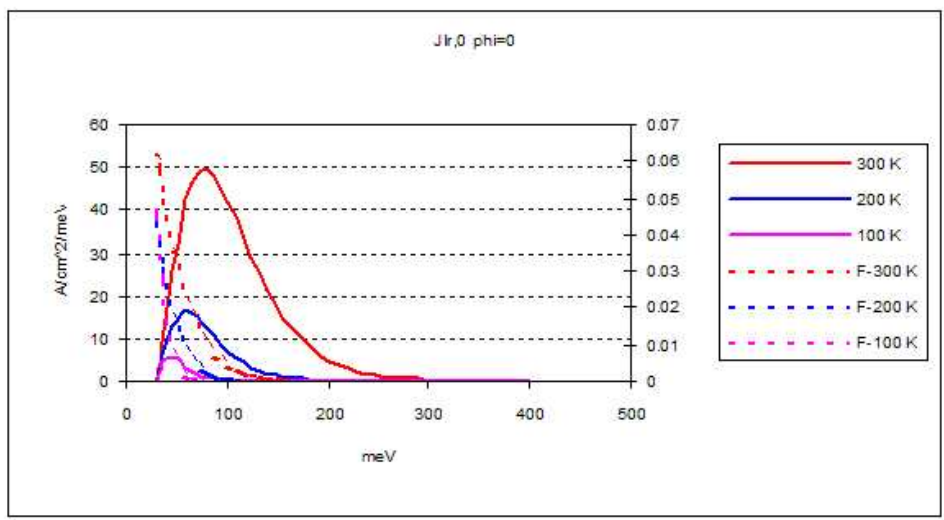

b)

Fig. 2 Diagrams and density of the current of elastic tunneling.

Fig.2 a) – tunneling of the free electron through the barrier from the left to the right. The value of the matrix element of this process is precisely equal to the reverse one, depicted in Fig. 2 b);

Diagrams of tunneling with consideration to the process of electron-phonon interaction of the first order are presented in Figs. 3 – 6. As will be shown below, in case of anisotropic hetero-structures their contribution into the full current going through the barrier will be asymmetrical.

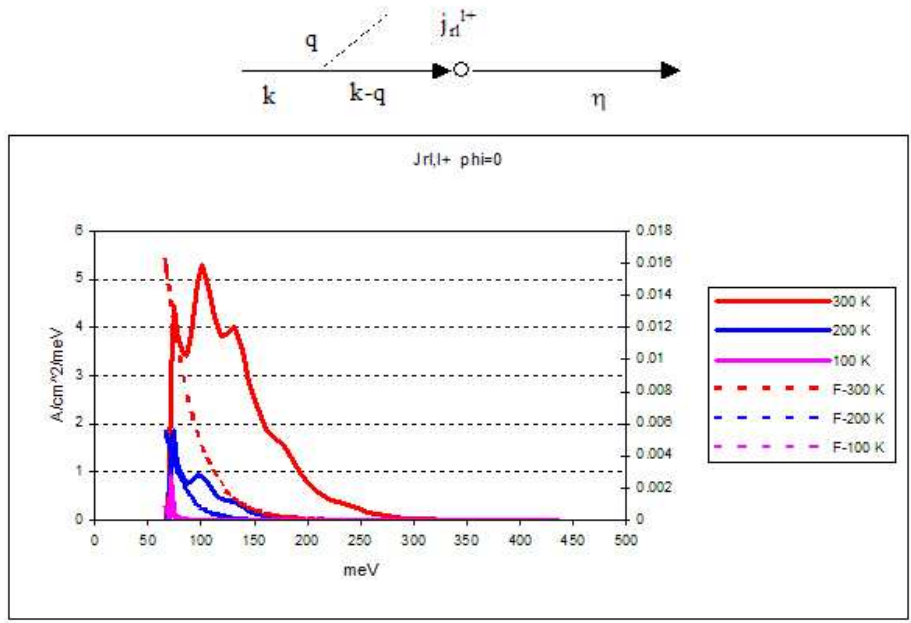

c)

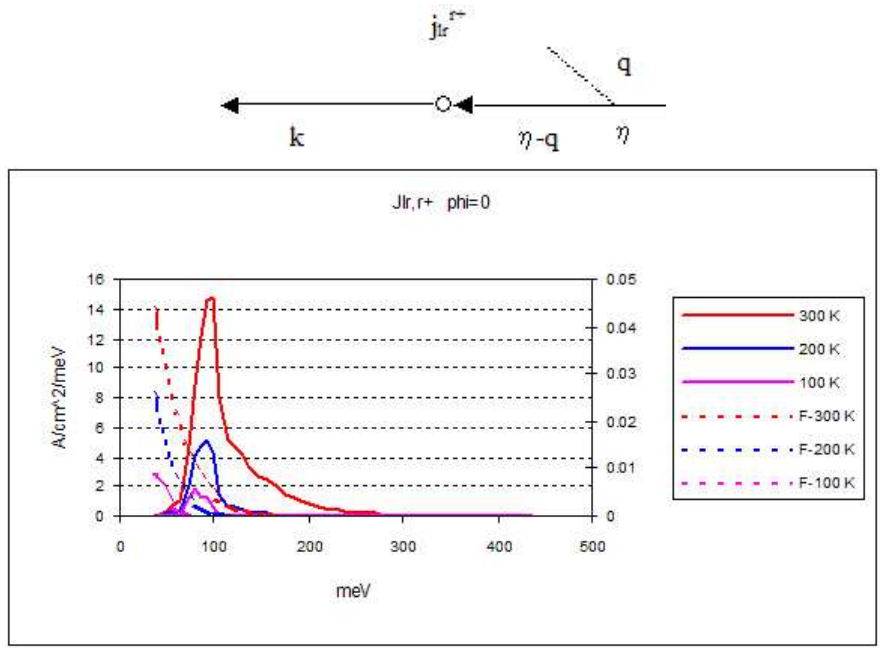

d)

Fig. 3 Diagrams and density of the current of inelastic tunneling

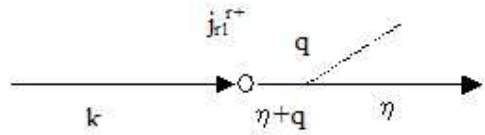

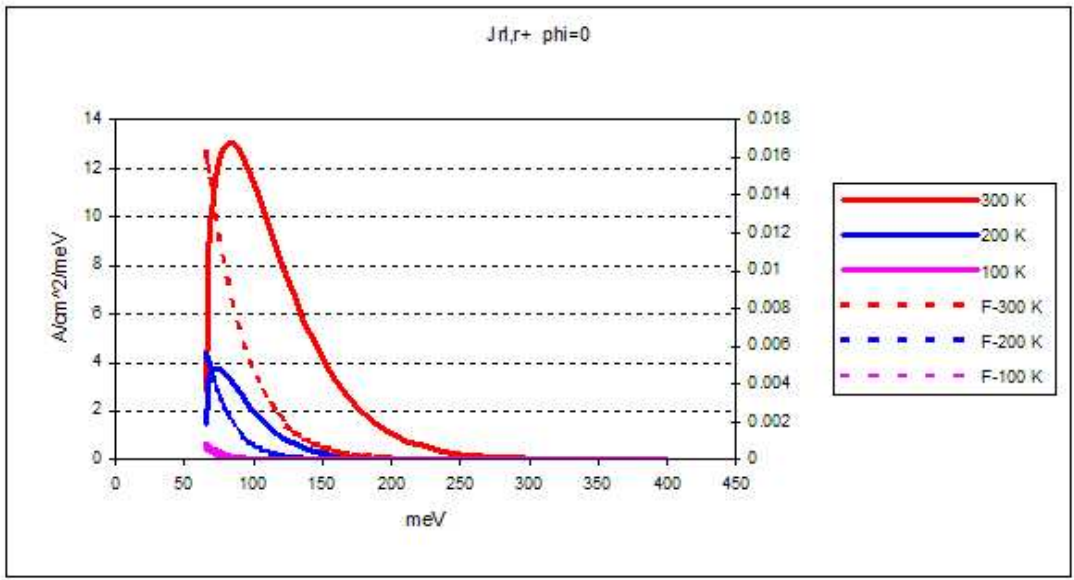

e)

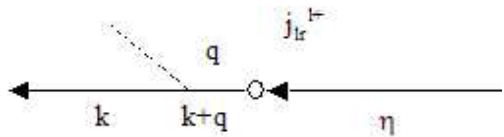

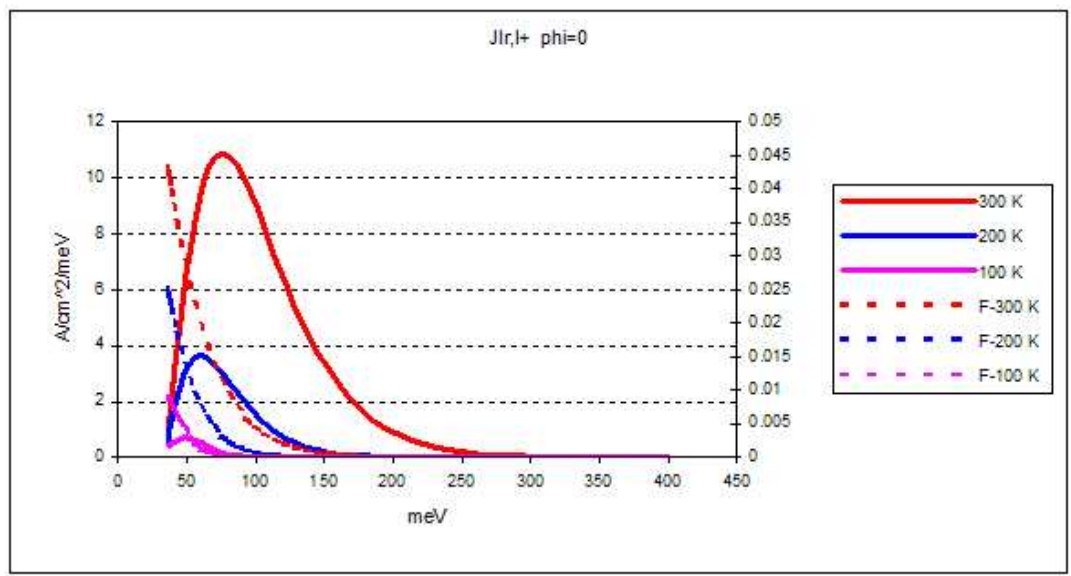

f)

Fig. 4 Diagrams and density of the current of inelastic tunneling

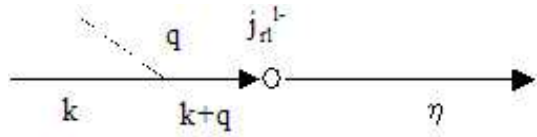

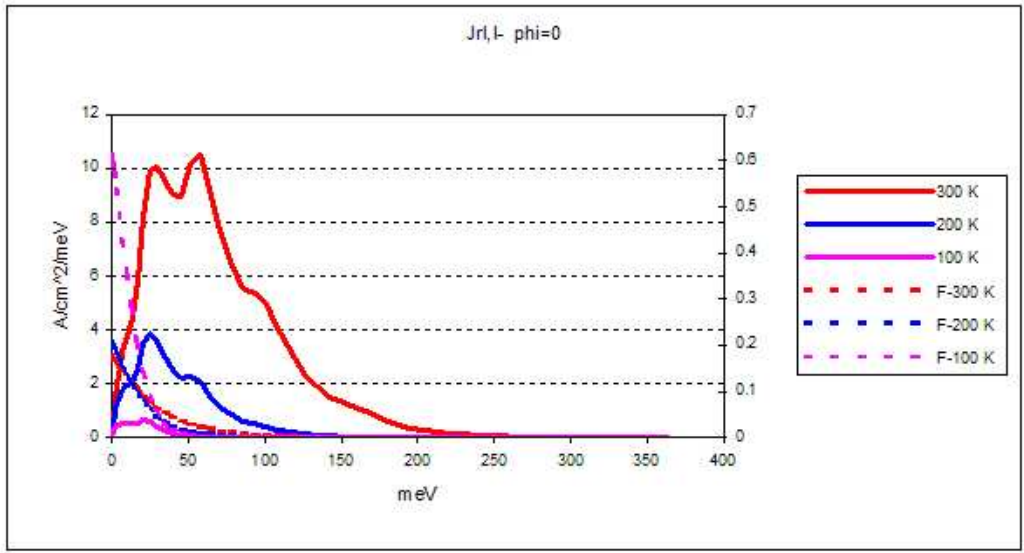

g)

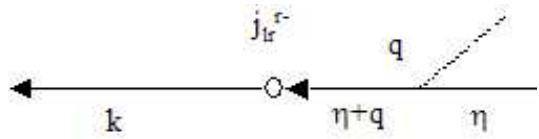

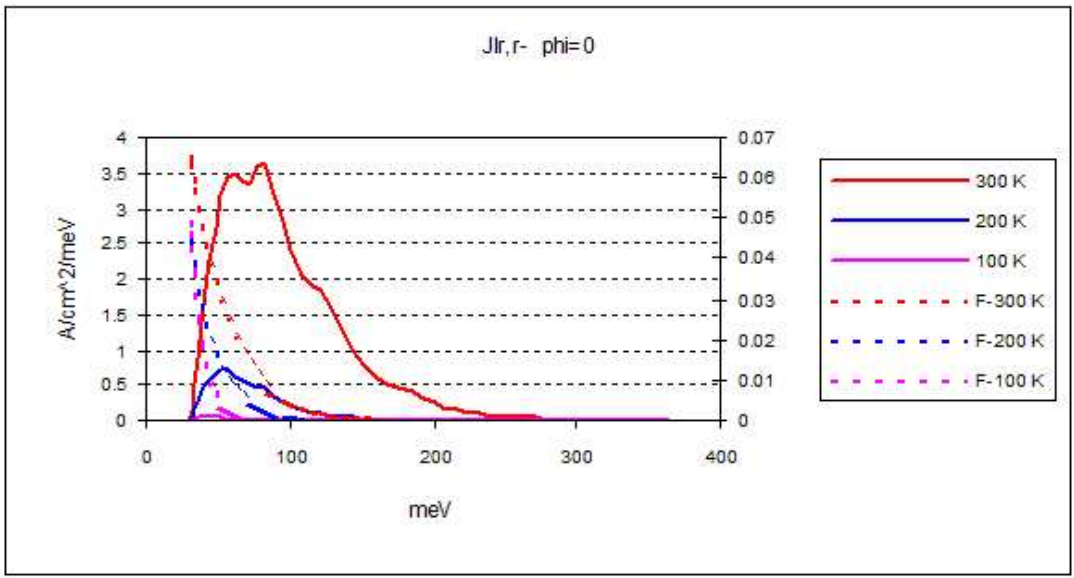

h)

Fig. 5 Diagrams and density of the current of inelastic tunneling

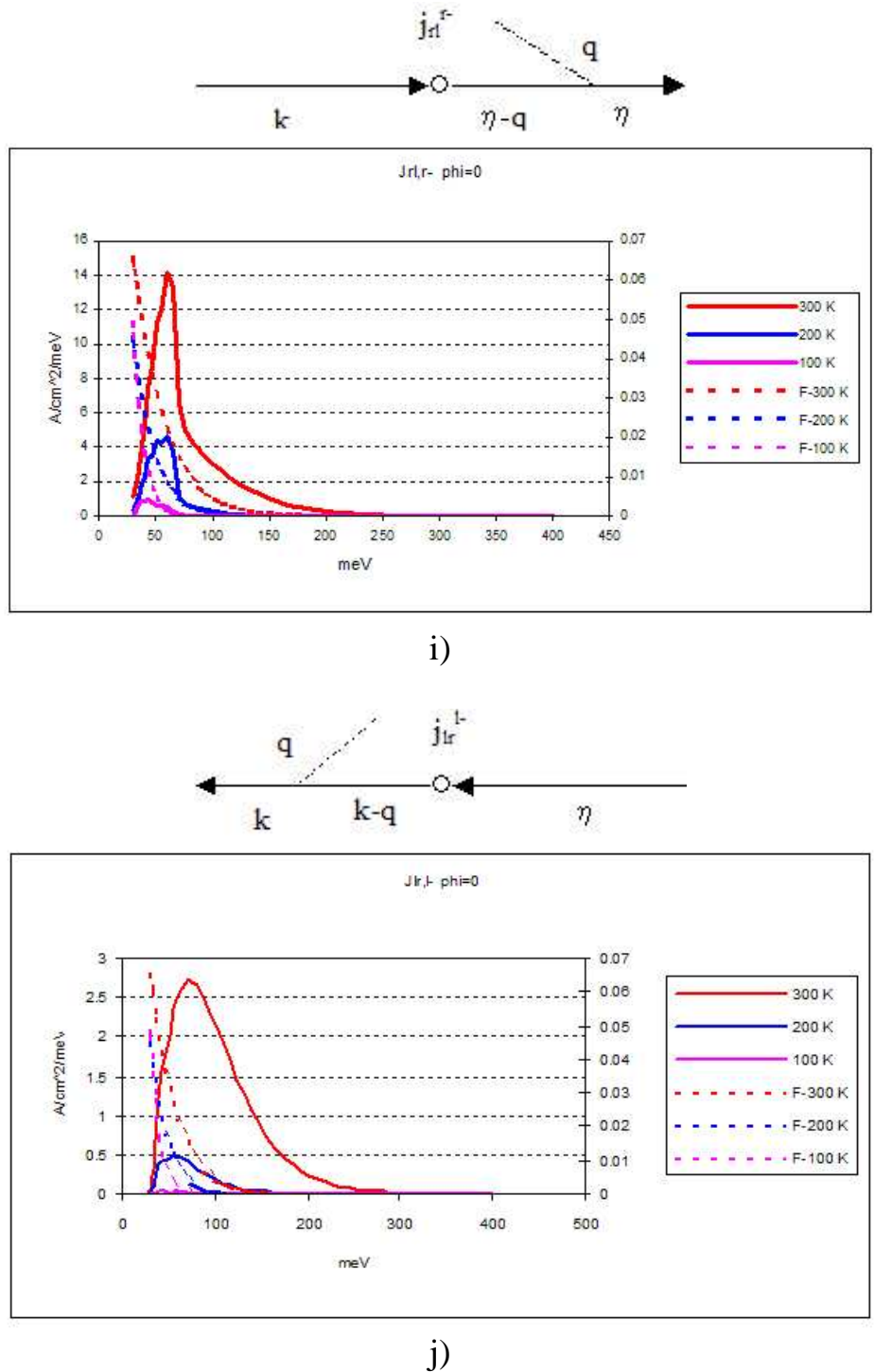

Fig. 6 Diagrams and density of the current of inelastic tunneling

***Asymmetry of phonon spectra of the adjacent to the barrier media.***

It should be noted, that the tunnel barrier demonstrates high transparency to acoustic waves, which, as it seems, has to result in similar character of oscillations of the lattice in the adjacent regions on both sides of the barrier, from which electrons are released. However, it is quite evident, at first, that if the adjacent to the barrier regions are physically different, then the oscillations in them can not be identical, no matter how close they are. This means, that the matter is not the    case, if a similarity exists here or not, but how strong this similarity is to be recorded.

Further, such oscillation modes may exist, which propagate only in one of the medium and exponentially attenuate in another. Let us consider as an example two linear chains, consisting from two kinds of atoms. Due to simulative character of the example all the parameters will be presented in relative units. Let the correlation of masses in one chain correspond to GaAs, and in another – to AlAs. Thus, $m_{Ga}=70$, $m_{Al}=30$, $m_{As}=75$. Frequency of optical oscillations for the wave vector q=0 in GaAs chain $\omega_{GaAs}(0)=36.25$, and in AlAs chain $\omega_{AlAs}(0)=50.09$. Frequency of optical oscillations at the boundary of region near AlAs will be 42,33, whereas near GsAs it will be 26.07. Dispersion curves for these chains are presented in fig. 7.

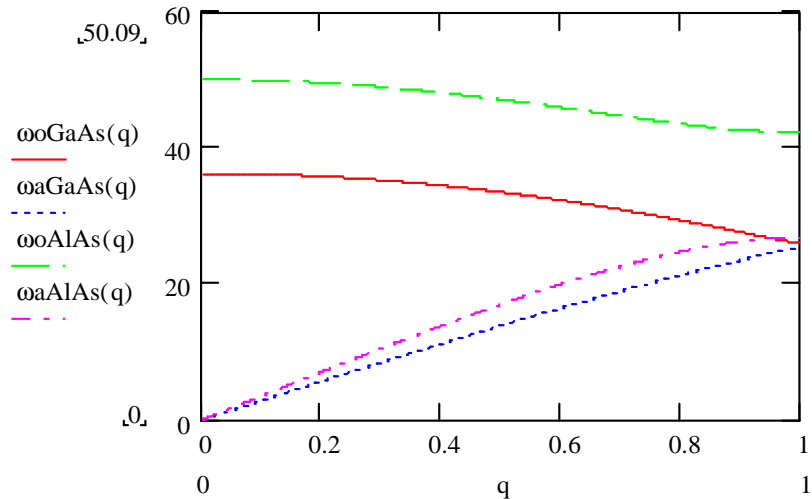

Fig. 7

In fig. 7 the dispersion curves for acoustic and optical modes of oscillations of simulative one-dimensional two-atom chains of atoms are presented.

Thus, it can be seen, that in this case all the modes of optical oscillations for one chain can not propagate in another chain, but will rather attenuate there. In general it is understandable, that there can be some oscillation modes (or phonons in a certain energy range), which may be general for both the barrier and the two media. However, such phonons should exist, which can live only on the left or the right side of the barrier.

Above all, even general phonons should have different wavelengths on different sides of the barrier. That is why their interaction with the wave function of electron can not be similar. Finally, electrons, possessing similar energy, on different sides of the barrier will have different wavelengths, thus interacting with phonons differently.

### Barriers with quantum dots or the impurity levels.

Asymmetry of transparency can be observed in barriers with quantum dots or the impurity levels (see /46/, for instance). Let the potential pit (quantum dot) in the barrier be such, that at least one quasi-stationary state of electron exists in it. Let the lifetime of electron in the potential pit be more than the time of dielectric relaxation of the environ-ment, the pit located asymmetrically to the middle of the barrier, for instance closer to its left boundary. With relaxation absent, i.e. at the independent from time profile of the potential barrier, probability of transition of electron of conductivity, for instance, from the right semi-space to the localized state in the pit and the reverse transition from the pit to the right semi-space will be equal. The same has to be true for electron exchange of the pit with the left semi-space. However for the case under consideration – the asymmetrically located potential pit, frequency of exchange of electrons between the pit and the left semi-space will be higher than that of exchange in the reverse direction. It is easy to demonstrate, that the equal to zero full current going through the barrier will settle at the equal Fermi levels of the adjacent to barrier media.

Let us remember at this point, that during the lifetime of electron in the pit the medium manages to relax. The pit in this case becomes deeper and the full energy pf electron drops. As a result of energy drop the velocities of the reverse transitions of electrons from the pit into semi-spaces will become lower than the velocities of the direct transitions. In the case under consideration, due to different widths of the barriers, separating the pit from electrically conductive semi-spaces, the velocity of transition from the pit into the right semi-space has to decrease stronger, than that of the left one. Here the current equal to zero, going through the barrier, settles already at the different Fermi levels for the adjacent to the barrier media. In our case of the so-called "preferable" transition of electron through the quantum dot from the right to the left, the equal to zero current has to settle at the decreased Fermi level of the right medium, as relates to the left one. It becomes evident than, that if we abridge these two media to one another without barriers, then in the circle formed the current equal to zero has to be sustained in a stationary mode.

### Calculations of current density in approximation of tunneling Hamiltonian with consideration the electron-phonon interaction.

Generally speaking, the subsequent consideration of tunneling asymmetry has to be done by methods of density matrix or Green function. However, in some approximation, in case of relatively low frequency of electron dissipation on another quasi-particles, and even lower frequency of tunneling through the barrier, the situation can be explained by inserting the vectors of states of electrons in the contacting media "l" and "r".

According to the abovementioned qualitative analysis, and as it will be shown in more detail further, the matrix elements of the tunneling Hamiltonian $|T_{lr}|^2$ in general case, will not be obligatory equal to $|T_{rl}|^2$. That is why the more refined as compared to (1) from /18, page 65/ expression for current j will look like:

$$j = \frac{4\pi e}{\hbar} \sum_{k_t} \int dE N_l N_r \left[ |T_{lr}|^2 f_l(1-f_r) - |T_{rl}|^2 f_r(1-f_l) \right], \qquad (1)$$

where e – elementary charge; $\hbar$ - Planck constant; $N_l$, $N_r$ и $f_l$, $f_m$ –densities of sates and distribution functions for media "l" and "r", correspondingly; $k_t$ – transverse wave vector; E – energy of electron (hole).

As can be easily seen from (1), at $|T_{lr}|^2 \neq |T_{rl}|^2$, the stationary state, at which j=0, settles with unequal Fermi levels of media $F_l$ and $F_r$.

Let us show in more detail how    anisotropy of tunneling transparency of the barrier occurs in case of virtual processes of absorption and radiation of phonons by the tunneling electron in the adjacent to the barrier physically different media. Let us consider, that the temperature, included into distribution functions of Fermi and Bose, is equal all around the system.

To solve the problem it is convenient to use the method of tunneling Hamiltonian, first used to describe tunneling process in solid bodies by Bardin /43/ and Harrison /44/. We assume here, that the lifetime of quasi-electron tends to infinity. This gives us the ability to approximate spectral function of electron /19, page 140/ δ - function of difference between energy and square of pulse of electron, which significantly simplifies integration of the expressions.

The following dependence of electron energy in the left and the right media adjacent to the barrier takes place:

$$E_l = \frac{\hbar^2 \cdot \left(k_z^2 + k_t^2\right)}{2 \cdot m_l} + U_{cl} \qquad (2)$$

$$E_r = \frac{\hbar^2 \cdot \left(\eta_z^2 + \eta_t^2\right)}{2 \cdot m_r} + U_{cr}, \qquad (3)$$

where $k_z$, $k_t$, $\eta_z$, $\eta_t$, - longitudinal and transverse wave vectors; $m_l$, $m_r$ – effective masses of electron; $U_{cl}$, $U_{cr}$ – levels of bottom of the conductive band.

The standard method of tunneling Hamiltonian /18, pages 13, 19, 30/ gives the ability to identify only diagonal elements of Hamiltonian matrix, interfacing the states with equal energy. However, to consider the contribution of radiation and absorption of phonons into the tunneling current (depicted as diagrams in figs. 3-6), and to consider the corresponding virtual states, its needs to specify both diagonal and non-diagonal components of the matrix of tunneling Hamiltonian, interfacing the states with different energy. To do this, we have to solve the problem in the following setup.

Let the potential energy U(z) prior to time t = 0 have the character of a step (see fig. 8 (b) below).

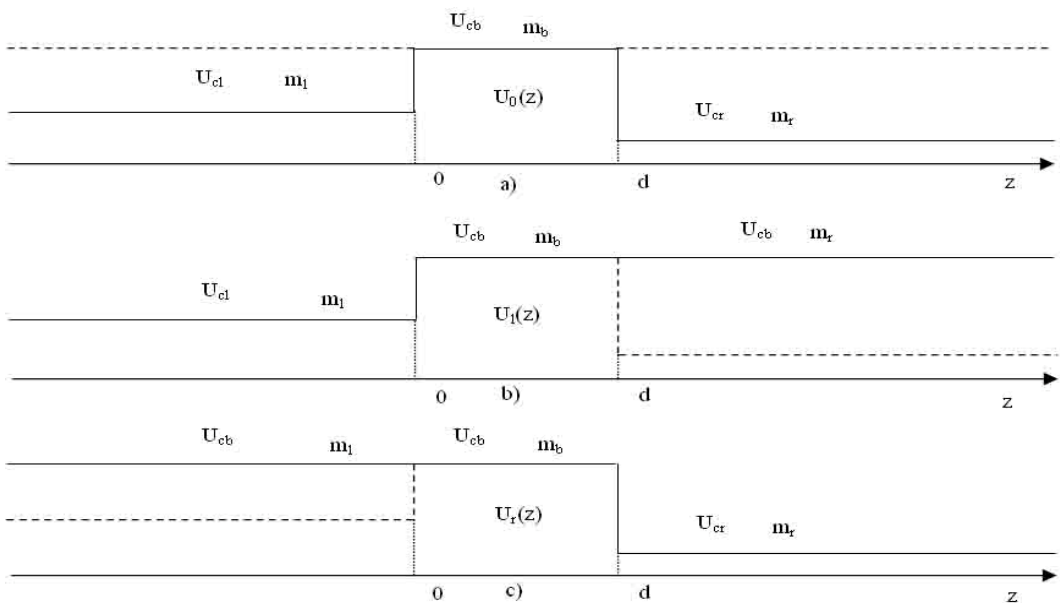

Fig. 8

Shown in fig. 8 are the profiles of the potential topography, used for calculation of the tunneling Hamiltonian.

Fig. 8a) is a picture of rectangular barrier $U_0(z)$, for which the tunneling Hamiltonian is calculated. $U_{cb}$, $U_{cl}$, $U_{cr}$ – are the bottom levels for conductive bands of the barrier and the adjacent media; $m_b$, $m_l$, $m_r$ - are the effective masses of electron in the barrier and the adjacent media.

Fig. 8b) is the picture of the initial potential $U_l(z)$, corresponding to the initial state of electron on the left of the barrier.

Fig. 8c) is a picture of potential $U_r(z)$, for which the functions of the final state of electron, tunneling from the left to the right were calculated.

$$U_l(z) = \Theta(-z) \cdot U_{cl} + \Theta(z) \cdot U_{cb} \qquad (4)$$

The initial state of electron is described by the wave function $\Psi_{lk}(t,x)$, having the form of a standing wave at $z < 0$ and exponentially attenuating at $z > 0$. When wave functions are cross-liked at the boundaries of the barrier, the difference of the effective masses of electron in the barrier is to be considered, as well as those on the left and on the right of the same.

The perturbing potential $V(t, x)$ effects the system being considered

$$V(t,\overline{x}) = -\left(U_{cb} - U_{cr}\right) \cdot \Theta(t) \cdot \Theta(z-d) \qquad (5)$$

Then, for $t > 0$ the potential relief takes the form of the barrier $U_0(z)$. To simplify calculations, the precise wave functions, corresponding to the potential shown in fig. 8a), and describing the electron on the right of the barrier, is approximated by the proper function $\Psi_{rn}(t,x)$ of the potential from fig. 8c).

The conditions of the task in the present setup allow to use the methods of non-stationary theory of perturbations, which offers the ability to correctly define the speeds of transitions. To simplify the record, we will use the standard manner of consideration of wave function in the finite volume (with barrier area S), then direct this volume to infinity and further pass from sums to integrals. The wave functions will be written as:

$$\Psi(t,\overline{x}) = a_{\overline{k}}(t) \cdot \Psi_{l\overline{k}}(t,\overline{x}) + \sum_{\overline{\eta}}\left[b_{\overline{\eta}\overline{k}}(t) \cdot \Psi_{r\overline{\eta}}(t,\overline{x})\right] \qquad (6)$$

Initial values of coefficients will be defined as follows: $a_k(0) = 1$; $b_{\eta k}(0) = 0$; $\dot{a}_k(0) = 0$. The task is to specify $\dot{b}_{\eta k}(0)$. Having substituted (6) into Schredinger equation, we will have, in a standard way, the following expression:

$$\dot{b}_{\overline{\eta}\overline{k}} = -\frac{i}{\hbar}\int d^3x \Psi^*_{r\overline{\eta}} \cdot V \cdot \Psi_{l\overline{k}} \qquad (7)$$

Making use of (7), and standard methods from (for instance. /42, c. 183/), the velocity of elastic transition of electron through the barrier from the left to the right $\dot{P}^0_{rl}$ looks like:

$$\dot{P}^0_{rl} = \frac{2\pi}{\hbar} \cdot \left|T^0_{rl}\right|^2 \cdot S \cdot (2\pi)^2 \cdot \delta^2\left(\overline{\eta}_{rt} - \overline{k}_{lt}\right) \cdot \delta\left(E_r - E_l\right) \qquad (8)$$

Expression (8) considers, in its explicit form, energy conservation and the transverse pulse of electron. The process is described by the diagram presented in fig. 2 a.

In this case the velocity of back transitions (diagram from fig. 2 b) turns to be equal to that of direct transitions $\dot{P}_{rl}^0 = \dot{P}_{lr}^0$ (evidently, in this case equality of matrix elements $\left|T_{rl}^0\right|^2 = \left|T_{lr}^0\right|^2$ can be stated). The explicit expression for the matrix element of the tunneling Hamiltonian has the following form:

$$T_{rl}^0 = \frac{4\hbar^2 m_b k_{lz} \eta_{rz} \chi_{r2} \cdot \exp(-\chi_{r2} d)}{\sqrt{\left(m_b^2 k_{lz}^2 + m_l^2 \chi_{l2}^2\right) \cdot \left(m_b^2 \eta_{rz}^2 + m_r^2 \chi_{r2}^2\right)}} \cdot \frac{(m_l \chi_{l2} + m_b \chi_{r1})}{(m_l \chi_{r2} + m_b \chi_{r1})} , \tag{9}$$

where $\qquad k_{lz} = \sqrt{\dfrac{2m_l}{\hbar^2}(E_{lk} - U_{cl}) - k_t^2}$ , $\qquad \eta_{rz} = \sqrt{\dfrac{2m_r}{\hbar^2}(E_{r\eta} - U_{cr}) - \eta_t^2}$ ,

$\chi_{r2} = \sqrt{\dfrac{2m_b}{\hbar^2} \cdot (U_{cb} - E_{r\eta}) + \eta_t^2}$ , $\chi_{l2} = \sqrt{\dfrac{2m_b}{\hbar^2} \cdot (U_b - U_r) + k_t^2}$ , $\chi_{r1} = \sqrt{\dfrac{2m_l}{\hbar^2} \cdot (U_{cb} - E_{r\eta}) + \eta_t^2}$ .

It is easily demonstrated, that in case of equality of the effective masses, as well as of the initial and final energy, the matrix element (9) corresponds to the well-known expression /18, с. 14/.

Making use of (8) and having integrated it for momentum space, we will have the dependence of the current going through the barrier from the left to the right $j_{rl}^0$ о on Fermi levels $F_l$ и $F_r$ :

$$j_{rl}^0 = \frac{1}{S} 2 \int \frac{d^3 k}{(2\pi)^3} 2 \int \frac{d^3 \eta}{(2\pi)^3} f_l(E_{lk}) \cdot \left[1 - f_r(E_{r\eta})\right] \cdot \dot{P}_{rl}^0 , \tag{10}$$

where $f_l$ и $f_r$ – functions of Fermi distribution.

The similar expression can be used for current $j_{lr}^0$ going through the barrier from the right to the left. The full current through the barrier in its zero approximation as to electron-phonon interaction, is equal to the difference $j_n^0 = j_{rl}^0 - j_{lr}^0$. Current $j_n^0 = 0$ at $F_l = F_r$.

Let us find the correction for the density of tunneling current, stipulated by electron-phonon interaction. Hamiltonian H for the whole system can be written as:

$$\hat{H} = \hat{H}_{0l} + \hat{H}_{0r} + \hat{H}_{lef} + \hat{H}_{ref} + \hat{H}_t , \tag{11}$$

where $H_{0l}$, $H_{0r}$, $H_{lef}$, $H_{ref}$, $H_t$ – Hamiltonians of non-interacting electron and phonon sub-systems, Hamiltonian of electron-phonon interaction and the tunneling Hamiltonian, correspondingly.

$$\hat{H}_{0l} = \int \frac{d^3 k}{(2\pi)^3} E_{lk} b_k^+ b_k + \int_{Dl} \frac{d^3 q}{(2\pi)^3} \hbar \omega_{lq} a_q^+ a_q \tag{12}$$

$$\hat{H}_{0r} = \int \frac{d^3 \eta}{(2\pi)^3} E_{r\eta} d_k^+ d_k + \int_{Dr} \frac{d^3 \nu}{(2\pi)^3} \hbar \omega_{r\nu} c_\nu^+ c_\nu \tag{13}$$

$$\hat{H}_{lef} = \int \frac{d^3 k}{(2\pi)^3} \int_{Dl} \frac{d^3 q}{(2\pi)^3} v_{lq} b_{k+q}^+ b_k (a_q - a_{-q}^+) \tag{14}$$

$$\hat{H}_{ref} = \int \frac{d^3 \eta}{(2\pi)^3} \int_{Dr} \frac{d^3 \nu}{(2\pi)^3} v_{r\nu} d_{\eta+\nu}^+ d_\eta (c_\nu - c_{-\nu}^+) \tag{15}$$

$$\hat{H}_t = \int \frac{d^3k}{(2\pi)^3} \int \frac{d^3\eta}{(2\pi)^3} \left[ T_{rl}(\eta,k) d_\eta^+ b_k + T_{lr}(k,\eta) b_k^+ d_\eta \right], \tag{16}$$

where $b^+$, $b$, $d^+$, $d$, $a^+$, $a$, $c^+$, $c$ – operators of creation and annihilation of electrons and phonons on the left and on the right of the barrier. Integration for wave vectors of phonons is done within Debye spheres $D_l$ and $D_r$, $v_{lq}$ and $v_{rv}$ - potentials of electron-phonon interaction.

Tunneling velocity, in the first order of theory of perturbations, was essentially found above (expressions 8-10). Let us find the corrections of the second order to vectors of electron states $\left| \Phi(t) \right\rangle$, including the effects of both the tunneling Hamiltonian and the Hamiltonian of electron-phonon interaction on the initial vector. The following expression can be taken as an example:

$$\left| \Phi(t) \right\rangle = \left( \frac{-i}{\hbar} \right)^2 \int_0^t dt_1 \int_0^{t_1} dt_2 \hat{H}_t(t_1) \hat{H}_{lef}(t_2) \left| k \right\rangle \tag{17}$$

This expression describes the processes, shown in diagrams from fig. 3a) and fig. 5a). Using the standard methods we can deduce the expressions for the velocities (per the unit of contact area) of the processes like $\dot{P}_{rl}^{l+}(\eta,k)$, including phonon radiation in the left semi-space, transition into the virtual state with the wave vector **k-q**, followed by tunneling into the finite state with the wave vector **η**:

$$\dot{P}_{rl}^{l+}(\overline{\eta},\overline{k}) = \frac{2\pi}{\hbar} \frac{\left| T_{rl}(\eta,k-q) \right|^2 \left| v_{lq} \right|^2}{\left( E_{lk} - E_{lk-q} - \hbar\omega_{lq} \right)^2} \delta\left( E_{lk} - E_{r\eta} - \hbar\omega_{lq} \right) \cdot (2\pi)^2 \delta^2 \left[ \overline{\eta}_t - \left( \overline{k} - \overline{q} \right)_t \right] \tag{18}$$

It is essential, that in the present case only the transverse pulse is preserved alongside with energy. The longitudinal pulse, due to interaction of electron with the barrier, is not preserved. This circumstance significantly changes the conditions, in which phonon radiation is studied, as compared to the case, when electron moves in homogeneous medium.

Making use of (18) we will have the constituent of the current density, stipulated by the processes, looking like the ones presented in fig. 3 a:

$$j_{rl}^{l+} = 2 \int \frac{d^3k}{(2\pi)^3} \int \frac{d^3\eta}{(2\pi)^3} \int_{Dl} \frac{d^3q}{(2\pi)^3} f_l(E_{lk}) \cdot \left[ 1 - f_r(E_{r\eta}) \right] \cdot \left[ 1 + g(\hbar\omega_{lq}) \right] \cdot \dot{P}_{rl}^{l+}(\eta,k), \tag{19}$$

where g – the function of Bose distribution.

The complete correction to tunneling current, stipulated by the processes of electron-phonon interaction of the first order, depicted in figs. 3-6, can be written as the sum of expressions like (18), (19) as:

$$j_n^1 = j_{rl}^{l+} + j_{rl}^{l-} + j_{rl}^{r+} + j_{rl}^{r-} - j_{lr}^{r+} - j_{lr}^{r-} - j_{lr}^{l+} - j_{lr}^{l-}, \tag{20}$$

Evidently, in this case the symmetry, characteristic of $j_n^0$ is absent.

Let us consider, for instance, the process, presented as diagram from fig. 3a). Electron on the right of the barrier with the initial wave vector **k** (and energy $E_{lk}$) radiates the phonon with energy $\hbar\omega_{lq}$, transits into virtual state with energy $E_{lk-q}$, and tunnels into the finite state on the right of the barrier with energy $E_r$. The reverse in time process is depicted as diagram from fig. 6b). However, the current component $j_{rl}^{l+}$ differs from $j_{lr}^{l-}$ since in the first case the integral has multiplier $\left[1 + g(\hbar\omega_{lq})\right]$, whereas in the second case it is $g(\hbar\omega_{lq})$. The difference is obviously related to the fact, that in the first case radiation of phonon is possible, into "phonon vacuum" including, whereas for the second case to be realized the presence of "phonon gas" is required. Above all, due to the difference of the effective masses $m_l$ и $m_r$ on the left and on the right of the barrier the matrix elements $T_{\eta,k\text{-}q}$ и $T_{k\text{-}q,\eta}$, at unequal initial and final energies become different, which can be proved by integration (7). The process, similar to the one depicted in Fig. 3a) is the process from fig. 3b). However, due to different dispersion characteristics of phonons on both sides of the barrier, sub-integral multipliers $\left[1 + g(\hbar\omega_{lq})\right]$ and $\left[1 + g(\hbar\omega_{rq})\right]$ will also differ. Above all, the Debay spheres, according to which integration is performed, will not coincide; matrix elements $T_{\eta k\text{-}q}$ и $T_{k\eta\text{-}q}$ observed as different as well.

In fig. 9 the contribution of elastic and inelastic components into the density of tunneling current, going from the left to the right and from the right to the left though it, is depicted as well as the summarized density of tunneling current.

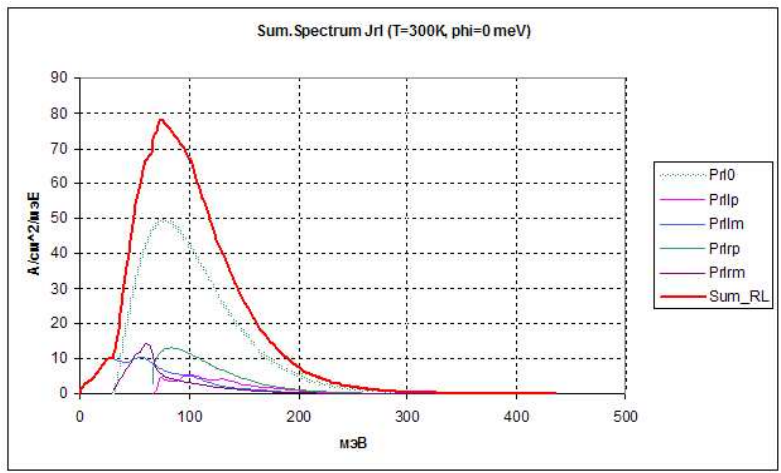

a)

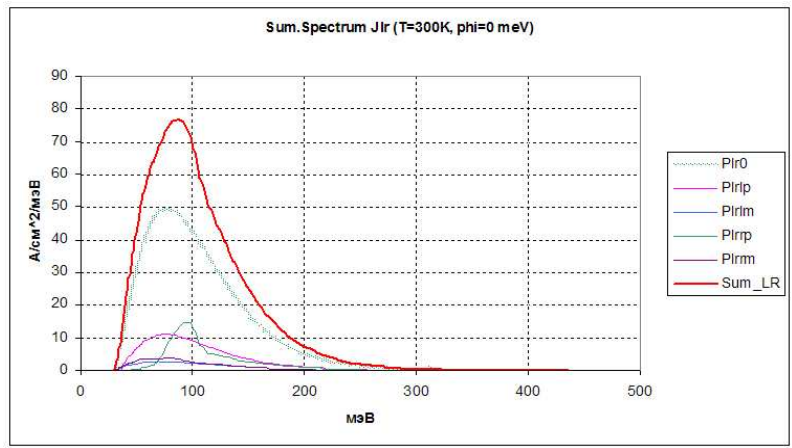

b)

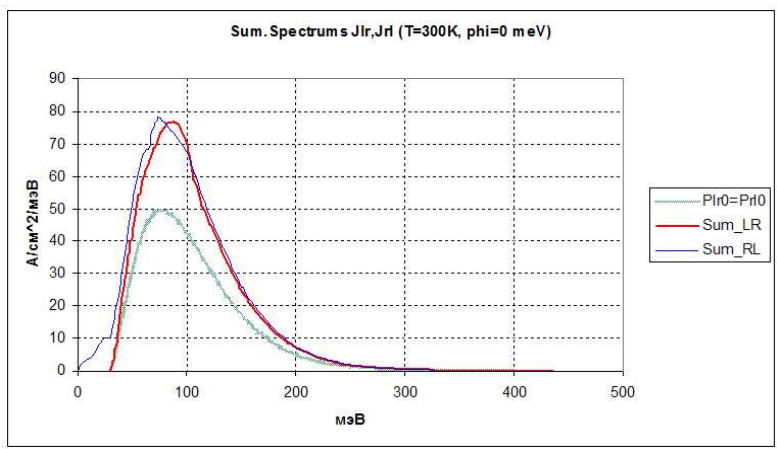

c)

Fig. 9

Thus, as opposed to tunneling of non-interacting electrons, electron-phonon interaction even of the first order, as considered in anisotropic hetero-structures, results in the tunneling through he barrier $j_n = j_n^0 + j_n^1$ turning to be equal to zero at $F_l=F_r$. Equality of currents is achieved at the shift of Fermi levels for some value $V_0$. As it follows from calculation results, presented in fig. 10, the value and the sign of $V_0$ depend on temperature.

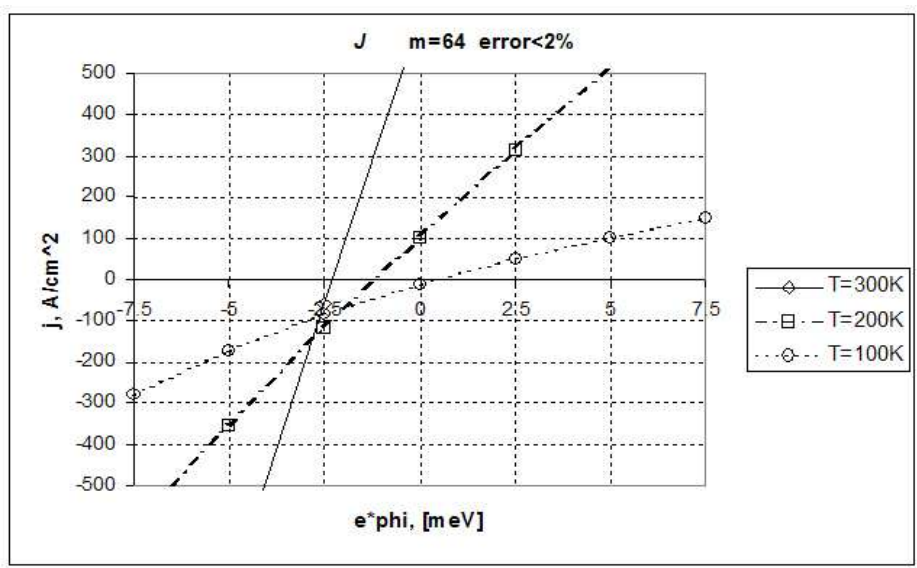

a)

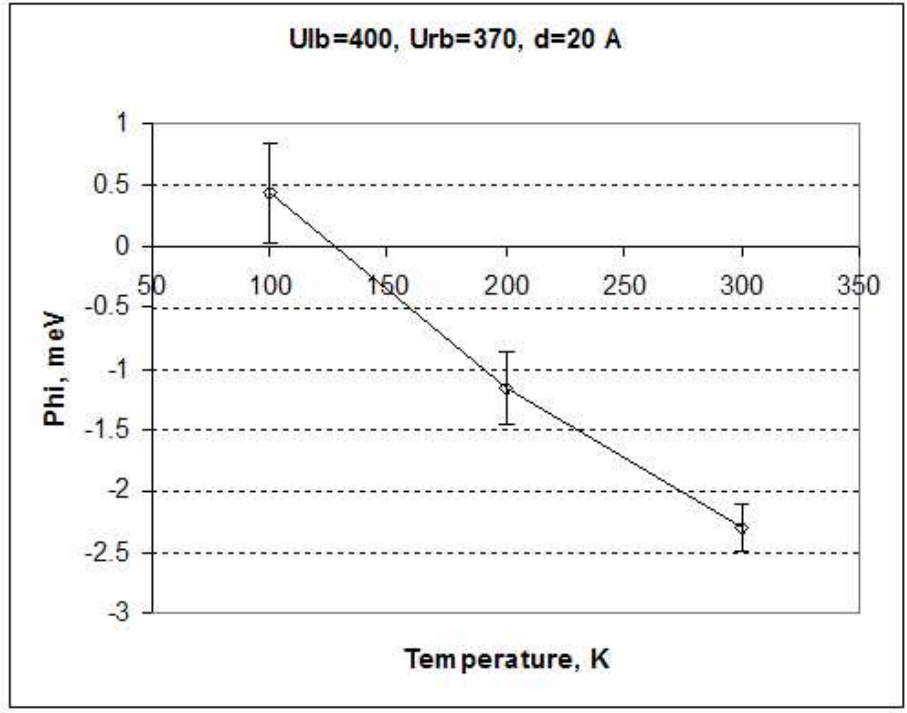

b)

Fig. 10

Fig. 10a) shows volt-ampere characteristics of GaAs/Al$_y$Ga$_{1-y}$As/Al$_x$Ga$_{1-x}$As. The level of replacement of gallium by aluminum in the barrier – y=0.36, replacement of gallium by aluminum in the right semi-space – x=0.03; in this case the height of the barrier is 400 meV, elevation of the conductive band on the right of the barrier is 30 meV, barrier thickness is 20 Angstrom. Calculated data are presented for temperature T = 100 K, 200 K and 300 K. In fig. 10b) the shift of Fermi levels (electromotive force) for the present structure versus temperature is presented.

***Estimation of tunneling current via quantum dots.***

Let us find how electron energy changes in a spherical potential pit, stipulated by polarization of lattice. Hamiltonian of electron interaction featuring wave function $\Psi_d(x)$ with the polarizing lattice can be written as:

$$\hat{H}_{ef} = \int d^3x \int d^3x_1 \frac{\rho_e(x) \cdot \hat{\rho}_f(x_1)}{|\bar{x} - \bar{x}_1|}, \qquad (21)$$

Electric discharge $\rho_f$, stipulated by polarization of lattice, can be expressed in a standard way (see, for instance /35, 36/) via operators of annihilation and creation of phonons a and a$^+$:

$$\hat{\rho}_f = -div\hat{P} \qquad (22)$$

$$\hat{P}_\mu = i \int \frac{d^3q}{(2\pi)^3} \sqrt{\frac{\hbar}{2\gamma_q \omega_q}} \cdot \left(a_{\mu\bar{q}} - a_{\mu-\bar{q}}^+\right) \cdot \exp(i\overline{qx}) \qquad (23)$$

$$\rho_e = -e|\Psi_d(\bar{x})|^2 \qquad (24)$$

$$\gamma_q = \frac{4\pi}{\omega_q^2} \cdot \left( \frac{1}{\varepsilon_\infty} - \frac{1}{\varepsilon_0} \right)^{-1}, \qquad (25)$$

where $\varepsilon_0$ и $\varepsilon_\infty$ are static and optical dielectric permittivity of medium.

Using (21-25) we have the following expression for the Hamiltonian $H_{ef}$:

$$\hat{H}_{ef} = -4\pi e \sum_\mu \int\limits_{Ld} \frac{d^3 q}{(2\pi)^3} \sqrt{\frac{\hbar}{2\gamma_q \omega_q}} \cdot \frac{q_\mu}{q^2} \cdot \left( a_{\mu \bar{q}} - a_{\mu -\bar{q}}^+ \right) \cdot \int d^3 x \left| \Psi_d(\bar{x}) \right|^2 \exp(i\bar{q}\bar{x}) \qquad (26)$$

Change of electron energy, stipulated by polarization of ion lattice (radiation and absorption of virtual phonons), can be found in the second order of theory of perturbations:

$$E_\Delta = -\int\limits_{Ld} \frac{d^3 q}{(2\pi)^3} \cdot \frac{\langle 0|\hat{H}_{ef}|q\rangle \langle q|\hat{H}_{ef}|0\rangle}{\hbar \omega_q} \qquad (27)$$

Let electron stay in a spherical potential pit with $r_0$ radius in the basic state, featuring energy $E_0$ (fig. 11). Then its wave function depends only on radius r and has the following form:

$$\Psi_0(r) = \frac{A}{r} \cdot \left\{ \theta(r_0 - r) \cdot \sin(kr) + \theta(r - r_0) \cdot \sin(kr_0) \cdot \exp[-\chi_0 \cdot (r - r_0)] \right\}, \qquad (28)$$

where A – normalization factor.

Parameters k и $\chi_0$ are defined by the expressions:

$$tg(kr_0) + \frac{k}{\chi} = 0 \qquad (29)$$

$$\chi_0 = \frac{\sqrt{2m_b \cdot (U_b - E_0)}}{\hbar} \qquad (30)$$

$$E_0 = \frac{\hbar^2 k^2}{2m_b} + U_0 \qquad (31)$$

It can be shown in this case, that:

$$E_\Delta \approx -\frac{e^2}{r_0} \cdot \left( \frac{1}{\varepsilon_\infty} - \frac{1}{\varepsilon_0} \right) \qquad (32)$$

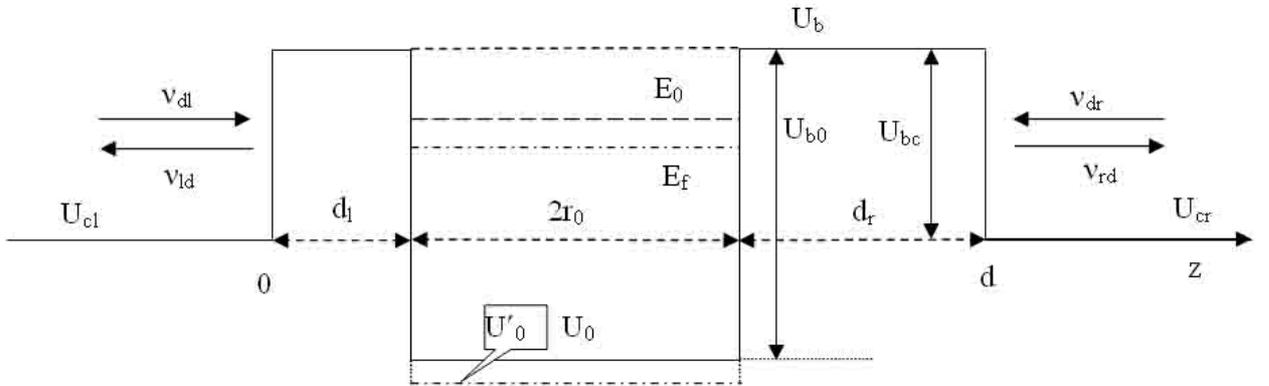

Fig.11

$E_0$ – the level of energy of the localized state of electron in the potential pit without medium relaxation taken into consideration; $E_f$ – the level of energy of the localized state of electron in the potential pit after medium relaxation; $\nu_{dl}$, $\nu_{ld}$, $\nu_{dr}$, $\nu_{rd}$ – frequencies of the direct and reverse electron transitions from the adjacent to the barrier media into quantum dot.

In fig. 11 the energy diagram of the barrier, containing the potential pit (quantum dot) is shown. The potential pit is assumed to be spherically symmetrical with $r_0$ radius. The minimum distance from the pit surface to the boundary of the left semi-space is $d_1$, to the right semi-space – $d_r$, $U_{cr}$, $U_0$, $U'_0$, $U_b$ – bottom levels of the conducting bands of the left and the right semi-spaces, the pits being in the initial relaxed state; the level of the bottom of the conductive band of the barrier $E_0$ is energy level of the localized state of electron in the potential pit, medium relaxation not considered for. $E_f$ is energy level of the localized state of electron in the potential pit after relaxation of medium. $\nu_{dl}$, $\nu_{ld}$, $\nu_{dr}$, $\nu_{rd}$ are frequencies of the direct and backward transitions of electron from the adjacent to the barrier media to the quantum dot.

Thus, in case of quite a big "lifetime" of electron in the localized state $\tau_d$, so, that the quantum uncertainty of energy $\Delta E \sim \hbar/\tau_d \ll |E_\Delta|$ and the bigger lattice relaxation time $(\sim 1/\omega_q)$, the difference of energies of localized state has to be observed, to which electron "comes" and from which the same "leaves" (energy levels $E_0$ and $E_f$, fig.11).

Let us consider the process of electron exchange of quantum dot with the adjacent to the barrier electrically conductive media. We will assume, that electron may be in the following states: a) on the left of the barrier, in state $\Psi_{lk}(x)$; b) on the right of the barrier in state $\Psi_{r\eta}(x)$; c) in the potential pit, in the initial state $\Psi_0(x)$ (without polarization ion lattice); d) in the potential pit, in the final state $\Psi_f(x)$ (after polarization of ion lattice by the field of polarized electron). In the last case it can be assumed, that the wave function of electron has the form (28), but parameter $\chi_0$ is to be exchanged for $\chi_f$:

$$\chi_f = \frac{\sqrt{2m_b \cdot (U_b - (E_0 + E_\Delta))}}{\hbar} \qquad (33)$$

Since $E_\Delta < 0$, then $\chi_f > \chi$. Correspondingly, the wave function of electron, staying in the deeper potential pit, attenuates quicker.

Following the method of tunneling Hamiltonian, the matrix elements of electron transition from the left to the right into the quantum dot - $T_{lk}$ and $T_{r\eta}$, correspondingly, can be determined, as well as those for electron escape from the quantum dot - $T_{kl}$ and $T_{\eta r}$. Obviously, in the first case we have to use functions $\Psi_{lk}(x)$, $\Psi_{r\eta}(x)$ and $\Psi_0(x)$, whereas in the second case - $\Psi_{lk}(x)$ $\Psi_{r\eta}(x)$ and $\Psi_f(x)$. For transition of electron from the left semi-space to the localized state on the quantum dot we will have:

$$T_{lk} = -\frac{\hbar^2}{2m_b} \int_{z=0} dS \left[ \Psi_0(r)\frac{\partial \Psi_{lk}(z)}{\partial z} - \Psi_{lk}(z)\frac{\partial \Psi_0(r)}{\partial z} \right] \cdot \exp(i\vec{k}, \vec{\rho}), \qquad (34)$$

where $\bar{\rho}$ - radius-vector in plane $z = 0$ or $z = d$.

Similarly the last matrix elements are determined:

$$T_{r\eta} = -\frac{\hbar^2}{2m_b} \int\limits_{z=d} dS \left[ \Psi_0(r) \frac{\partial \Psi_{r\eta}(z)}{\partial z} - \Psi_{r\eta}(z) \frac{\partial \Psi_0(r)}{\partial z} \right] \cdot \exp(i\overline{\eta}_t \overline{\rho}),$$ (35)

$$T_{kl} = \frac{\hbar^2}{2m_b} \int\limits_{z=0} dS \left[ \Psi_f(r) \frac{\partial \Psi_{lk}(z)}{\partial z} - \Psi_{lk}(z) \frac{\partial \Psi_f(r)}{\partial z} \right] \cdot \exp(-i\overline{k}_t \overline{\rho}),$$ (36)

$$T_{\eta r} = \frac{\hbar^2}{2m_b} \int\limits_{z=d} dS \left[ \Psi_f(r) \frac{\partial \Psi_{r\eta}(z)}{\partial z} - \Psi_{r\eta}(z) \frac{\partial \Psi_f(r)}{\partial z} \right] \cdot \exp(-i\overline{\eta}_t \overline{\rho}),$$ (37)

By using matrix elements (34-37) we can find the frequencies of electronic exchange of quantum dot with the adjacent to the barrier media. With the changed for $E_\Delta$ value electron energy in pre-exponential multipliers and distribution functions neglected, we will have:

$$\nu_{dl} = \nu_0 \cdot f_l \cdot (1 - f_d) \cdot \exp(-\alpha_l),$$ (38)

$$\nu_{ld} = \nu_0 \cdot f_d \cdot (1 - f_l) \cdot \exp(-\beta_l),$$ (39)

$$\nu_{dr} = \nu_0 \cdot f_r \cdot (1 - f_d) \cdot \exp(-\alpha_r),$$ (40)

$$\nu_{rd} = \nu_0 \cdot f_d \cdot (1 - f_r) \cdot \exp(-\beta_r),$$ (41)

where $\quad \nu_0 = \dfrac{(E_0 - U_0)}{4\pi\hbar} \cdot \left( \dfrac{E_0 - U_0}{U_b - U_0} \right)^{3/2}, \quad f_l = f_l(E_0), \quad f_r = f_r(E_0), \quad \alpha_l = 2\chi_0 \cdot d_l,$

$\beta_l = 2\chi_f \cdot d_l, \ \alpha_r = 2\chi_0 \cdot d_r, \ \beta_r = 2\chi_f \cdot d_r, \ f_d$ - probability of occupation of quantum dot.

In a stationary state the following equality has to be met:

$$\nu_{dl} + \nu_{dr} = \nu_{ld} + \nu_{rd},$$ (42)

The current through quantum dot is defined by the following differences:

$$j_1 = \nu_{dl} - \nu_{ld} = \nu_{rd} - \nu_{dr},$$ (43)

Making use of (42) one can find $f_d$

$$f_d = \frac{f_l \cdot \exp(-\alpha_l) + f_r \cdot \exp(-\alpha_r)}{f_l \cdot \exp(-\alpha_l) + f_r \cdot \exp(-\alpha_r) + (1 - f_l) \cdot \exp(-\beta_l) + (1 - f_r) \cdot \exp(-\beta_r)},$$ (44)

Let the surface density of quantum dot be equal to $N_{dS}$. Then the final expression for density of the current going through the barrier can be written as:

$$j_d = \nu_0 N_{dS} \frac{f_l \cdot (1 - f_r) \cdot \exp(-\alpha_l - \beta_r) - f_r \cdot (1 - f_l) \cdot \exp(-\alpha_r - \beta_l)}{f_l \cdot \exp(-\alpha_l) + f_r \cdot \exp(-\alpha_r) + (1 - f_l) \cdot \exp(-\beta_l) + (1 - f_r) \cdot \exp(-\beta_r)},$$ (45)

As it follows from (45), with relaxation of the lattice absent, when $\alpha_l = \beta_l$ and $\alpha_r = \beta_r$, the equal to zero current settles at $F_l = F_r$. However, at asymmetrically located quantum dots, when $d_l \neq d_r$ and with relaxation considered for, when $\alpha_l \neq \beta_l$ и $\alpha_r \neq \beta_r$, the equal to zero current $j_d$ settles at $F_l \neq F_r$.

The above processes can take place at the background of non-correlated with the electron under consideration "symmetric" thermal fluctuations of the barrier height, and the probable spatial technological inhomogeneity.

### Simulated example of bi-stable asymmetric barrier system.

One more example of presence of asymmetry of probability of backward transitions of particles through the barrier in anisotropic structures in classical systems can be made. Anisotropy of transparency may happen in the case, for instance, where the barrier is represented by bi-stable asymmetric system, for which the ability of smooth transition to one or another option of state depend on the initial conditions. Let the barrier be the system of two spatially separated ion layers of different mass and value of elastic forces. It can be expected, that the sample particle, moving from the ion layer with lower mass (higher elastic mobility), compresses the barriers, thus forming the high value of the maximum potential. As it moves from the ion layer with higher mass (lower mobility), the particle moves the barriers apart (moving apart the light and mobile ions), - in this case the barrier is lower in height. Should the critical parameter be exceeded, which, in this case, are particle coordinates, the jump transition into the alternative branch of states occurs.

Occurrence of anisotropy of transparency can be exemplified by the model of double barrier. Let the first barrier be immovably fastened in the origin of coordinates and have the interaction potential $U_a(x)$ with the incident particle.

$$U_a(x) = \frac{U_0}{\sqrt{\left(\frac{x}{a}\right)^2 + 1}} \qquad (46)$$

The second mobile barrier forms the potential $U_b(x,y)$

$$U_b(x,y) = \frac{U_0}{\sqrt{\left(\frac{x-y}{a}\right)^2 + 1}}, \qquad (47)$$

where y – the coordinate of the barrier tip.

Elastic properties of the barrier are characterized by the function of potential system energy versus coordinate y.

$$U_c(y) = U_{c0} \cdot \left(\frac{y}{d} - 1\right)^2, \qquad (48)$$

In expressions (46-48) values $U_0$, $U_{c0}$, a- are parameters, d – coordinate of the tip of mobile barrier in the absence of exterior effects.

Let a particle be approaching the system under consideration. Then its potential energy will be $U_{si}(x,y) = U_a(x) + U_b(x,y)$, and parameter i may take values 1 and 2, depending on the direction. Coordinate "y" can be found from the condition of maximum of summarized potential energy of the mobile barrier at the specified coordinate of particle "x" - $U_f(x,y) = U_b(x,y) + U_c(y)$. Simulating calculations were performed for parameters $U_0=1$, $U_{c0}=1$, $a=1$, $d=5$. Let us assume, that the particle approaches the barrier from the left side. Then the coordinate of the mobile barrier corresponds to the minimum of the right potential pit $U_f(x,y)$ (fig.12a). Here the particle potential corresponds to the curve $U_{s2}$ (fig. 12b). In point $x \approx 9.1$ the barrier collapses into the left potential pit, the potential of the particle dropping spasmodically. Maximum of the potential barrier $U_{max1} \approx 1.2$ is achieved in this case in point $x \sim 0$. When the particle approaches the barrier from the right, the mobile barrier is initially located in the left potential pit. The particle potential corresponds to the curve $U_{s1}$. In point $x \sim 9$ the barrier collapses into the right potential pit. Maximum of the potential barrier $U_{max2} \approx 1.6$ is reached in point $x \sim 0.9$.

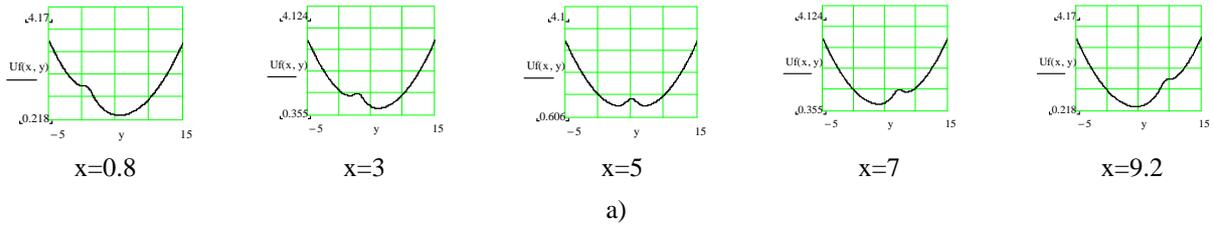

x=0.8    x=3    x=5    x=7    x=9.2

a)

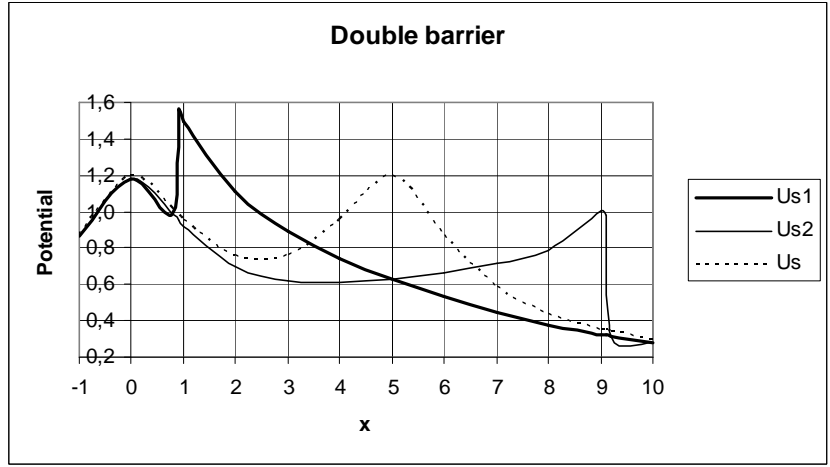

b)

## Fig. 12

Fig. 12 a) is the potential energy of mobile barrier versus the coordinate of the incident particle. Fig.12 b) presents the dependencies of the potential energy of the incident particle when it reaches the double barrier from the side of the mobile barrier (curve $U_{s1}$), immobile barrier (curve $U_{s2}$) and at the condition stating that both barriers are immobile (curve $U_s$).

Thus, the energy interval occurs, in which particles can move only in one direction. This illustrates probability of absence of reversibility of motion in the mechanical system incorporating classical particles.

### *Factors, stipulating asymmetry of micro-transitions.*

The following factors, stipulating asymmetry of micro-transitions though the barrier in anisotropic semi-conductor systems can be enumerated:

1. Asymmetric effect of correlating charges of the particle, crossing the potential barrier, on its height for direct and backward transitions, stipulated by the finite time of their relaxation;
2. Asymmetry of contribution into the summed current of inelastic tunneling processes, accompanied with radiation and absorption of quasi-particles and stipulated by the difference of spectral and statistic characteristics of the adjacent to the barrier bosonic sub-systems.
3. Anisotropic relaxing structure of the barrier per se.

It follows from the above, that strictly speaking, in any spatially inhomogeneous systems of interacting particles the precise thermodynamic equilibrium in stationary state, which is understood in terms of equality of medium values of thermodynamical parameters, is never achievable. In the majority of cases the difference between stationary and equilibrium states can be neglected. For instance, according to calculations, at character-istics values of parameters of the modern     Schottki diodes, featuring the required for occurrence of the effect anisotropy, stationary shift of Fermi levels, separated by metal barrier and semiconductor, is equal to practically not observable value ~ $10^{-20}$ eV. However, in media with specially formed structure this difference can produce observable effects, which may be even of commercial validity.

Essentially, hysteresis of the potential barrier follows from the well-known principle of Frank-Condone /27, page 372/, according to which electronic transitions are so fast, that the medium is not capable to trace them. This circumstance results, for instance, in the shift of spectral lines of absorption and radiation of F-centers in solid states. The close to the character relaxation effect takes place in ionization-recombination processes /28, page 105/, where at some certain parameters of plasma ionization of a neutral molecule occurs in the Coulomb field, whereas recombination – in the Debye one, which offers the corresponding corrections to the cross-sections of ionization and recombination. It is quite essential, that, as opposed to the above processes, asymmetry of transitions between micro-states is spatially directed in anisotropic hetero-structures during tunneling.

### Analysis of applicability of the second law of thermodynamics to the phenomena under consideration.

Since the phenomena under consideration tackle the second law of thermodynamics, their more profound statistic substantiation has to be given.

The basic assumption of statistic thermodynamics is the postulate of a-priori equiprobability: the closed system can be in any admissible quantum (microscopic) state with equal probability /5, page 31/. To follows, in particular, from this, that in equilibrium state the system has to possess maximum entropy as the logarithm of the number of micro-states. From the viewpoint of statistics it is more correctly to say, that some state, characterized by certain microscopic parameters (temperature, pressure, etc.) is observed more often and is accepted as the equilibrium one, because of the biggest number of micro-states corresponding to it.

Deviation of parameters from equilibrium values (fluctuations) correspond to lower number of admissible micro-states, and it is notable, that the less number of micro-states correspond to some fluctuation, the more rarely it is observed. It is worth mentioning, that in the basic assumption the trustworthy of the principle of detailed equilibrium is stipulated: only the numbers of micro-states are considered, but the intensity of transitions between them is concealed. Evolution of non-equilibrium systems, described by the kinetic equation (for gases, for instance), is accompanied with growth of their entropy /30, page 26/. It is essential, that to prove this assumption, known as H-theorem of Boltzmann, the principle of detailed equilibrium is used, expressing some certain symmetry of transitions in the system. In its turn, the above symmetry occurs as a result of disregard of correlation of particle distribution /31, page 173/.

Meanwhile, in the process of transition through the barrier a carrier interacts with the long-ranging field of the relaxing system (correlation coat), the state of which depends on pre-history of the process and, correspondingly, on the direction of motion and carrier energy. This makes the situation different from the simple processes of particle interaction – their dissipation, described by the symmetric with regard to reversal of time sign equations of classical and quantum mechanics.

The effect under consideration also occurs as a result of correlation of motion of a single microscopic particle (electron) taken into consideration, with all the entire ensemble of particles, forming the relaxing medium, undergoing evolution. Thus, the systems with asymmetric barriers are not covered with H-theorem of Boltzmann and the stationary state in them does not correspond to the equilibrium one.

Let us point, that figuratively speaking in the system under consideration anisotropic barrier plays the role of the so-called "Maxwell demon" /32. page 138/, "trapping" the particles, flying to one side, and letting them go, if they fly to another side. The problem, pronounced as far as in 1871, still remains acute /52/.

In /33/ the situation with "Maxwell demon" is analyzed and it is stated, that in the conditions of thermodynamic equilibrium of electromagnetic field (radiation) with a substance (particles) the "demon" will be blind, just as in the case, when radiation is absent at all. The cloud of light, coming from all the directions will take place, giving no indication of either dislocation or speed of the particles. So, the "demon" would not be able to let the particles through in a purposeful manner.

It is deemed, that in the analysis, mentioned above, some inaccuracies are present. Firstly, in spite of the equilibrium character of the averaged in time electromagnetic field, its fluctuations should be correlated with the motion of electrically charged particles. Consequently, the "demon" or some electronic device, playing its role, will not be absolutely blind in the conditions considered.

Secondly, the "demon" is apprehended as something exterior with respect to the medium, containing particles. Thus, to be ruled according to the exterior with respect to the medium laws, it has to receive information, process it and act somehow: change the exterior conditions in a purposeful manner. It was proved, however, that with relaxation in anisotropic structures considered for, the above "demonic" properties are the attribute of the medium per se. Asymmetry of transparency in this case occurs in compliance with the interior or the proper laws of particle motion in a medium. It follows, that in this specific case the "informational" substantiation of impossibility of the processes, accompanied with the decreasing entropy, is incorrect /34, page 398/.

The shift of thermodynamical parameters can be identified with a model system taken as example, consisting of two similar vessels of V volume containing N particles of the ideal gas and contacting via the transparent barrier. Permeability of the barrier is described by functions $\nu_{rl}$ и $\nu_{lr}$ – average frequencies of particle transitions with p pulse from the left to the right and vice versa. All the system is thermally insulated so, that its full energy 2E is preserved. Let us assume, that particle interchange is weak and gas density is low, i.e. distribution of particles as to their energy in each system follows the Boltzmann law. In case the frequencies $\nu_{rl}$ и $\nu_{lr}$ are equal, we have a standard task of the two systems with the diffusion contact between them (see, for instance /24, page 67/). Their stationary state, corresponding to maximum entropy, is characterized by equality of temperatures and chemical potentials (Fermi levels).

Being assumed is that $\nu_{rl} \neq \nu_{lr}$ . Then, the stationary state of the system shall be identified using the kinetic equations for the particle flux and energy flux through the barrier, provided that the full number of particles and full energy are preserved.

$$\dot{N}_l = -\dot{N}_r = \sum_p \left[ \nu_{lr}(p) \cdot f_{rp} \cdot \left(1 - f_{lp}\right) - \nu_{rl}(p) \cdot f_{lp} \cdot \left(1 - f_{rp}\right) \right] = 0 \qquad (49)$$

$$\dot{E}_l = -\dot{E}_r = \sum_p \left[ \nu_{lr}(p) \cdot \varepsilon_p \cdot f_{rp} \cdot (1 - f_{lp}) - \nu_{rl}(p) \cdot \varepsilon_p \cdot f_{lp} \cdot (1 - f_{rp}) \right] = 0 \qquad (50)$$

$$N_l + N_r = 2N \qquad (51)$$

$$E_l + E_r = 2E, \qquad (52)$$

where $N_i = \sum_p f_{ip}$, $E_i = \sum_p \varepsilon_p \cdot f_{ip}$, $f_{ip} = \exp\left( \dfrac{F_i - \varepsilon_p}{T_i} \right)$, $\varepsilon_p = \dfrac{p^2}{2m}$, m – mass of parti-cles, index i may take values 1 or 2.

Thus, the system, consisting of 4 equations results, used to identify 4 unknowns – $F_1$, $F_2$, $T_1$, $T_2$. The frequency of transitions can be presented as:

$$\nu_{rl}(p) = \nu_p + \gamma_p \qquad (53)$$

$$\nu_{lr}(p) = \nu_p - \gamma_p \qquad (54)$$

Then, provided that $f_{lp} \ll 1$, $f_{rp} \ll 1$, (49, 50), the following can be deduced:

$$\sum_p \nu_p \cdot (f_{rp} - f_{lp}) = \sum_p \gamma_p \cdot (f_{rp} + f_{lp}) \qquad (55)$$

$$\sum_p \nu_p \cdot \varepsilon_p \cdot (f_{rp} - f_{lp}) = \sum_p \gamma_p \cdot \varepsilon_p \cdot (f_{rp} + f_{lp}) \qquad (56)$$

Let us assume, that the parameters, characterizing, for instance, frequencies of transitions, are not dependant on particle energy, which in reality is never observed, but gives the illustrative and qualitatively proof results $\nu_p = \nu = \mathrm{const}$, $\gamma_p = \gamma = \mathrm{const}$. Then, taken (55, 56) as a basis, we have:

$$\nu \cdot (N_r - N_l) = \gamma \cdot (N_r + N_l) \qquad (57)$$

$$\nu \cdot (E_r - E_l) = \gamma \cdot (E_r + E_l) \qquad (58)$$

Thus, in the present case we have $N_r = N \cdot (1 + \gamma/\nu)$, $N_l = N \cdot (1 - \gamma/\nu)$, $T_l = T_r$, $S = S_0 - N \cdot (\gamma/\nu)^2$, where $S_0$ is entropy of the system at anisotropic transparency of the barrier. Anisotropic transparency of the barrier results in the reduced value of entropy of stationary state of the system, as compared to the maximum value, corresponding to thermodynamically equilibrium state, which is an attribute of isotropic barriers. The chaotic thermally-governed particle motion is self-organized into the directed motion. This circumstance makes it possible the effective work to be realized due to thermal energy of particles. It has to be pointed here, that in the present model, due to similar energy spectra of particle fluxes, overcoming the barrier, the temperatures of all the sub-systems turn to be equal.

The effect of anisotropic transparency of barriers will not contradict the second law of thermodynamics, if the wording of the last is itemized, the basic postulate of statistic thermodynamics included into the same, in particular as follows:

**"If a closed system can stay with the equal probability in any admissible microscopic state, and if the same system, at some moment of time, changes for non-equilibrium macroscopic state, then the most probable result in the upcoming moment of time will be the monotonous growth of entropy of this system".**

Then, due to asymmetry of frequencies of the contrary transitions of electrons through the barrier, the equal probability absent, as a consequence, the system under consideration turns to be outside the effect of the second law of thermodynamics in its clarified wording. Thus, in the processes being considered no contradictions with the above physical law are present, if the boundaries of its applicability are specified more strictly.

**The effect of barrier anisotropy: estimation of its observable parameters in a quantitative manner**.

To estimate the parameters, stipulated by anisotropy of the barrier in a quantitative manner, one has to write the expression for the current going through the barrier as a function of electrochemical potential, temperature and parameters of the lattice.

The following physical model will serve as the basis for selection of the option, used to implement the effect under consideration in practice and to identify its commercial applicability:

1. A closed circuit, consisting of electrically conductive medium, separated by the potential barrier and asymmetrically located quantum dots is present (see fig.11). For illustrative purposes the pictures of the medium are separated symbolically into 2 parts (("l" and "r") and are assumed to be cross-linked at their boundaries.

2. Equality of current $j_n$ to zero in the circuit in stationary state of the system is not assumed a-priori. It is only assumed, that the current in the barrier and media l and r is equal.

3. Temperature of T-ion and electron sub-systems is equal all over the circuit. Thus, rather high thermal conductivity of ion lattice is assumed as well as the comparatively low-intensive electron exchange of media through the barrier. In particular, it is being assumed, that each particle tunnels independently. Corrections to current density, stipulated by temperature difference of sub-systems with asymmetric micro-transitions, can be easily found.

4. To simplify calculations, non-degenerated semiconductors are considered, thus distribution of electrons as to energy in each of the media is assumed to be of Boltzmann character.

Let us specify the stationary state of plasma. Being assumed is that before they were locked into the loop through the barrier, the media contained $n_0$ of free electrons, correspondingly. Then, Poisson equation for electric potential $\varphi_\beta$ and the expression for particle flux density $j_n$, considering for the drift and diffusion components, will have the following form (index $\beta$ may take values "l" or "r"):

$$\frac{\partial^2 \varphi_\beta}{\partial z^2} = \frac{4\pi e}{\varepsilon_o}\left(n_\beta - n_o\right) \tag{59}$$

$$j_n = \mu \cdot n_\beta \frac{\partial \varphi_\beta}{\partial z} - D \cdot \frac{\partial n_\beta}{\partial z}, \tag{60}$$

where $\varepsilon_o$, $\mu$ D – static dielectric permeability, mobility and diffusion coefficient, correspondingly; $n_\beta$ - density of electrons.

Making use of (59, 60) the coordinate dependence for the bottom of conductive bands and Fermi energy can be found:

$$U_{cl}\left(z\right) = U_{cl}\left(-d\right) - \gamma \cdot \left(z+d\right) - C \cdot \left[\exp(\frac{z+d}{L_D}) - 1\right] \tag{61}$$

$$U_{cr}\left(z\right) = U_{cr}\left(d\right) - \gamma \cdot \left(z-d\right) + C \cdot \left[\exp\left(-\frac{z-d}{L_D}\right) - 1\right] \tag{62}$$

$$E_{Fl}(z) = E_{Fl}(-d) - \gamma \cdot (z + d) \tag{63}$$

$$E_{Fr}(z) = E_{Fr}(d) - \gamma \cdot (z - d), \tag{64}$$

where $\gamma = \dfrac{e \cdot j_n}{\mu \cdot n_0}$, $L_D = \sqrt{\dfrac{\varepsilon_0 \cdot T}{4\pi e^2 n_0}}$ - Debye screening length, $\mu$ - mobility, C- constant.

Density $n_0$ can be written as:

$$n_0 = N_C \cdot \exp\left(-\frac{\xi_0}{T}\right), \tag{65}$$

where $N_C$ – density of the number of states, $\xi_0$ – chemical potential.

Then, the density of mobile electrons versus coordinates will look like:

$$n_\beta(z) = n_0 \cdot \exp\left(\frac{\xi_0 - U_{c\beta}(z) + E_{F\beta}}{T}\right) \tag{66}$$

To determine the stationary state of the system 6 constants are to be identified: $j_n$, $E_{Fl}(0)$, $E_{Fr}(d)$, $U_{cl}(0)$, $U_{cr}(d)$, C. This can be done, using the boundary conditions (see fig. 13), the condition of preservation of electric charge and expression (45) for the current going through quantum dots. Density of electric charge inside the barrier is being neglected.

$$j_d\left(E_{Fl}(0), E_{Fr}(d), U_{cl}(0), U_{cr}(d)\right) = j_n \tag{67}$$

$$E_{Fr}(L) - U_{cr}(L) = \xi_0 \tag{68}$$

$$E_{Fl}(-L) = E_{Fr}(L) \tag{69}$$

$$U_{cl}(-L) = U_{cr}(L) \tag{70}$$

$$U_{cr}(d) = U_{cl}(-d) + U'_{cl}(-d) \cdot \frac{\varepsilon_0}{\varepsilon_{b0}} \cdot 2d \tag{71}$$

$$U_{cr}(L) = 0 \tag{72}$$

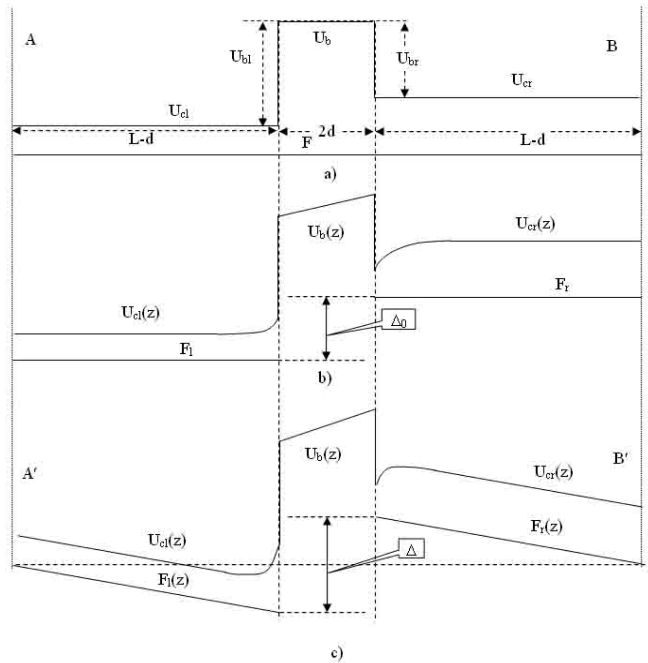

Fig.13

In fig. 13 energy diagrams for hetero-structures with the tunnel barrier are depicted.

In fig. 13a) a hetero-structure with the barrier, featuring isotropic transparency is presented with the specially selected (equal to F) Fermi levels of all the media, which stipulated the absence of distortion of the bottom of the conductive band in the vicinity of the barrier.

In fig. 13b) a hetero-structure with the barrier, transparency of which for tunneling of electrons is higher from the left to the right, than in the opposite direction, is presented. The rear surfaces of the adjacent to the barrier media (planes A-A' and B-B') are assumed to be electrically disconnected. Stationary state in this case (equal to zero current through the barrier) is achieved, when Fermi levels of the adjacent to the barrier media $F_l$ and $F_r$ shift for $\Delta_0$ value.

Fig. 13c) presents a hetero-structure with the barrier, transparency of which for electron tunneling is again higher from the right to the left than in the opposite direction. However, the rear surfaces of the adjacent to the barrier media (planes A-A' and B-B') are assumed to be electrically connected in this case. The stationary state here is achieved when Fermi levels of the adjacent to the barrier media $F_l$ and $F_r$ shift for $\Delta$ value and at the current, going through the barrier, which is not equal to zero.

From the viewpoint of electrical engineering, the anisotropic barrier exhibits itself so, that the volt-ampere characteristic of such a structure does not go through a zero point. This means, that this structure can be characterized with the interior resistance $R_0$ and the shift of electric potential (electromotive force) $V_0$. It can be written from the above:

$$j_d = \frac{1}{R_0} \cdot \left( U + V_0 \right) \tag{73}$$

$$eU = E_{Fl}\left(-d\right) - E_{Fr}\left(d\right) \tag{74}$$

With the system of equations (67-72) solved and with consideration for current (73) the current in the circuit will look like:

$$j_n = \frac{V_0}{R + \dfrac{L}{\mu \cdot n_0}} \tag{75}$$

Maximum energy release in the load (electrically conductive medium) $P_m$ is achieved at the condition (76) observed:

$$\mu \cdot n_0 \cdot R_0 = L \tag{76}$$

$$P_m = \frac{V_0^2}{2R_0} \tag{77}$$

### Estimation of current in the barrier with quantum dots.

Let us consider the device based on heterojunction structure GaAs – $Al_xGa_{1-x}As$ with quantum dots InAs as the first variant of technical implementation of the effect of anisotropic transparency of the barrier. The parameters of the materials involved are taken from /38, 47/.

Being assumed is that T = 300 K. The levels of gallium substitution with aluminum in the barrier x = 0.2, the height of the barrier being $U_{bc}$ = 200 meV. Also assumed is that one of the quasi-stationary levels of electron in quantum dot $E_0$ (fig.11) does not ex-ceed the level $U_c$ for more than kT value    (where  k  is  Boltzmann  constant).
Quantum dots are located in the barrier asymmetrically: $d_l = 40$ Å, $d_r = 50$ Å. Density of
the carriers in the adjacent to the barrier media is $n_0 \sim 10^{17}$ cm$^{-3}$. It follows then, that $E_\Delta \approx$
5meV, $\nu_0 \approx 10^{15}$ s$^{-1}$, $\chi_0 \approx 6.52 \cdot 10^6$ cm$^{-1}$, $\chi_f \approx 6.61 \cdot 10^6$ cm$^{-1}$, $\alpha_l \approx 5.22$, $\beta_l \approx 5.29$, $\alpha_r \approx 6.52$,
$\beta_r \approx 6.61$. Density of quantum dots per unit of area is $N_{ds} = 10^{11}$ cm$^{-2}$.

For the closed circuit the calculated electromotive force (stationary shift of Fermi
levels at the barrier) is $U_0 = 0.39$ mV, $R_0 = 1.4 \cdot 10^{-5}$ Ohm·cm$^2$, current $j = 14$ A/cm$^2$. The
maximum energy-conversion efficiency in this case per one layer of the structure of the
transformer is $P_m \approx 5$ mW/cm$^2$. Having assumed, that the thickness of one layer of the
transformer $H \approx 5 \cdot 10^{-6}$ cm, the specific energy-conversion efficiency will be $Q = P_m/H \approx 1$
kW/cm$^3$.

### Estimation of current in anisotropic hetero-structure.

Let us consider anisotropic structure GaAs/Al$_y$Ga$_{1-y}$As/Al$_x$Ga$_{1-x}$As as the second
option for implementation of the effect of anisotropic transparency. The level of substitu-
tion of gallium with aluminum in the right semi-space is x = 0.03, the height the barrier
being 400 meV, elevation of the conductive band on the right of the barrier being 30
meV, the thickness of the barrier 20 Angstrom.

As it follows from calculation results, presented in fig. 10, the effect of inelastic
tunneling, at the condition of asymmetry of phonon spectra, results, at temperature T =
300 K, in electromotive force at the barrier $V_0 = 2,5$ mV, the specific resistance of the
barrier being $R_0 = 4 \cdot 10^{-6}$ Ohm·cm$^2$. In this case the maximum energy-conversion effi-
ciency per one layer of the transformer structure will be $P_m \approx 3$ W/cm$^2$.

### Application options for the effect of anisotropic transparency.

The structures, featuring the commercially valuable energy-transformation proper-
ties, can be made of the materials, base on semiconductor heterojunction structure; con-
tacts like metal-semiconductor of the locking type; structures like metal-dielectric-semi-
conductor, as well as from another inhomogeneous media. In this case both mono-polar
and bi-polar (electron or hole) exchange of media with carriers can be used.

The technology, used to produce energy transformers, implies basically the conse-
quent application of the layers of semiconductor and metal materials on a substrate. This
can be done by thermo-vacuum, ion-plasma or magnetron deposition /39, page 290/ or by
methods of gaseous or liquid-phase epitaxy /40, page 503/.

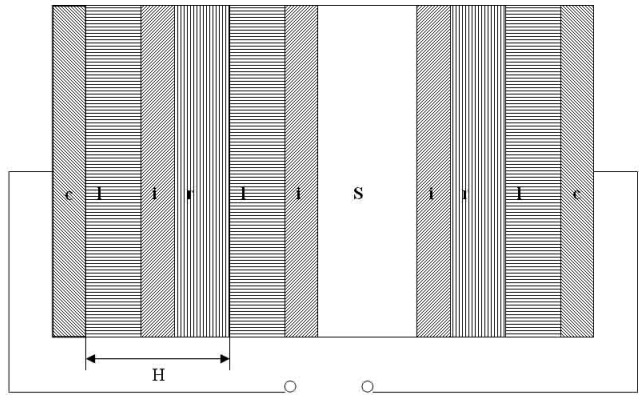

Fig. 14

In fig. 14 a schematic variant of the      device, based on the effect of barrier anisotropy is depicted. Symbols "l" and "r" stand for electrically conductive physically different layers of semiconductor (differing by correlation energy and (or) time of relaxation of carriers). Symbol "i" stands for tunnel-transparent barrier at heterojunction structure. Symbol "c" stands for the layer of metal, providing the ohm contact; H and S are thickness and area of cross-section per one layer of the transformer.

Due to high specific energy-conversion efficiency the area of commercial application of the effect under consideration can be very wide: from the inbuilt supply sources to ecologically friendly fuel-free transportation means and power stations. The effect of temperature decrease of the operating body of the transformer can be of a separate interest and can become topical in view of development of cooling system of various types, the ones, used for the elements of computerized machinery including.

*Allied problems.*

Study of possibility of processes with asymmetric micro-transitions, resulting in regulation and self-organization of particles in space under the natural conditions is of substantial scientific and technological interest.

Examination of non-local effect of correlating charges on the motion of charged particles in the relaxing system without barriers, but with the gradient of parameters, somehow essential for formation of correlation charges, is the endeavor, close to the topic of this paper. Inhomogeneous plasma can be taken as example. In this case the particles, staying in one plane, and having the equal but oppositely directed velocity along the gradient of medium parameter, should feature different energy due to the differently relaxing correlation charges.

## Literature cited.

1. Ф. Блатт. Физика электронной проводимости в твердых телах. М., "Мир", 1971.
2. В.Л. Бонч - Бруевич, С.Г. Калашников. Физика полупроводников. М., "Наука", 1977.
3. Большая советская энциклопедия. т. 4, с.599, М., "Советская энциклопедия", 1971.
4. К. А. Путилов. Термодинамика. М., "Наука", 1971.
5. Ч. Киттель. Статистическая термодинамика, М., "Наука", 1977.
6. Л.Д. Ландау и Е.М. Лифшиц. Статистическая физика. Часть 1, М., "Наука", 1976.
7. А.С. Зильберглейт, Г. В. Скорняков. Преобразование тепла в работу с помощью потенциальных систем. ЖТФ, т.62, в.2, с.190-195, 1992 г.
8. G.Krause. Patent DE, A1, 2929949 from 24.07.79, H02 N 11/00.
9. G.Krause. Patent DE, A1, 2950704 from 17.12.79, H02 N 11/00.
10. G.Krause. Patent т DE, A1, 2913730 from 16.10.80, H02 N 11/00.
11. M. Gourdine. Patent WO,A1, 82/00922 from 05.09.80, H01 L 35/00, 35/02, 35/28, 37/00; H01 M 2/38, 6/36, 14/00.
12. O. Bschorr. Patent DE, A1, 3315960 from 02.05.83, H01 L 35/26.
13. G., J. - E. Masse, K. Djessas. Patent FR, A1, 2680279 from 09.08.91, H01L 35/00; G01K 7/02; G01J 1/42.

14. Л.Д. Ландау и Е.М. Лифшиц. Теория упругости. М., "Наука", 1987.
15. В.М. Бродянский. Вечный двигатель - прежде и теперь. М., "Энергоатомиздат", 1989.
16. И.П. Базаров. Термодинамика. М., "Высшая школа", 1991.
17. С.Г. Лазарев. Влияние релаксации полупроводника на высоту барьеров Шоттки. ВАНТ, сер. "Теор. и прикл. физ.", в. 3, с. 3 (1996).
18. Туннельные явления в твёрдых телах. Под ред. Э. Бурштейна и С. Лундквиста. М., "Мир", 1973.
19. Е.Л. Вольф. Принципы электронной туннельной микроскопии. К., "Наукова Думка", 1990.
20. Л.Д. Ландау и Е.М. Лифшиц. Электродинамика сплошных сред. М., "Наука", 1982.
21. L.G. Ilchenco, E.A. Pashitskii and Yu. A. Romanov. Charge interaction in layered systems with spatial dispersion. Surf. Sci., v.121, p. 375 (1982).
22. С.М. Зи. Физика полупроводниковых приборов. Кн. 1. М., "Мир", 1984.
23. A.M. Gabovich, V.M. Rosenbaum and A.I. Voitenco. Dynamical image forces in three-layer systems and field emission. Surf. Sci., v.186, p. 523 (1987).
24. B.N. Persson and A. Baratoff. Self-consistent dynamic image potential in tunneling. Phys. Rev. B, v.38, N.14, p.9616 (1988).
25. K.L. Sebastian and G. Doyen. Dynamical image interaction in scanning tunneling microscopy. Phys. Rev. B, v.47, N.12, p.7634 (1993).
26. Н.Л. Чуприков. Туннелирование в одномерной системе N одинаковых потенциальных барьеров. ФТП, т. 30, в. 3, с. 443-453 (1996).
27. У. Харрисон. Теория твёрдого тела. М., "Мир", 1972.
28. Л.П. Кудрин. Статистическая физика плазмы. М., "Атомиздат", 1974.
29. Л.И. Глазман, Р.И. Шехтер. Неупругое резонансное туне-лирование электронов через потенциальный барьер. Т. 94, в. 1, с. 292-305 (1988).
30. Е.М. Лифшиц, Л.П. Питаевский. Физическая кинетика., М., "Наука",1979.
31. Г. Рёпке. Неравновесная статистическая механика. М., "Мир", 1990.
32. А.И. Ансельм. Основы статистической физики и термодинамики. М., "Наука", 1973.
33. Н. Винер. Кибернетика и общество. М., "Иностранная литература", 1958.
34. Р. Л. Стратонович. Теория информации. М., "Сов. радио", 1975.
35. Ч. Киттель. Квантовая теория твёрдых тел. М., "Наука", 1967.
36. Х. Хакен. Квантовополевая теория твёрдого тела. М., "Наука", 1980.
37. Поляроны. Под ред. Ю.А. Фирсова. М., "Наука", 1975.
38. S. Adachi. GaAs, AlAs, and $Al_xGa_{1-x}As$: Material parameters for use in research and device applications. J. Appl. Phys. Vol.58, numb.3, p.p. R1-R29, 1985.
39. В.Н. Черняев. Технология производства интегральных микросхем и микропроцессоров. М., "Радио и связь", 1987.
40. Справочник по электротехническим материалам. т. 3, под ред. Ю. В. Корицкого, В.В. Пасынкова, Б.М. Тареева, Л., "Энергоатомиздат", 1988.
41. А.С. Давыдов. Квантовая механика. М., Физматгиз, 1963.
42. Л.Д. Ландау и Е.М. Лифшиц. Квантовая механика. Нерелятивистская теория. М., "Наука", 1974.

43. J. Bardeen. Tunneling from a Many-       Particle Point of View. Phys. Rev. Let. Vol. 6, Numb. 2, pp. 57-59, 1960.

44. W. A. Harrison. Tunneling from an Independent-Particle Point of View. Phys. Rev. Vol. 123, Numb. 1, pp. 85-89, 1961.

45. С.И. Пекар. Исследования по электронной теории кристаллов. М.-Л., Гос. изд. тех. теор. лит., 1951.

46. T. Suzuki, Y. Haga, K. Nomoto, K. Taira and I. Hase. Tunneling current through self-assembled InAs quantum dots embedded in symmetric and asymmetric AlGaAs barriers. Solid State Electronics, Vol. 42, No. 7-8, pp. 1303-1307, 1998.

47. E. Herbert Li. Material parameters of InGaAsP and InAlGaAs systems for use in quantum well structures at low and room temperatures. Physica E 5 (2000), pp. 215-273.

48. Ю.Л. Болотин, А.В. Тур, В.В. Яновский. Нелинейное трение как механизм генерации направленных движений. ЖТФ, т. 72, в. 7, с. 9-12 (2002).

49. В.Л. Попов. Наномашины: общий подход к индуцированию направленного движения на атомном уровне. ЖТФ, т. 72, в. 11, с. 52-63 (2002).

50. Г.В. Скорняков. Преобразование тепла в работу с помощью термически неоднородных систем (исправление). Письма в ЖТФ, т. 23, № 5, с. 91-95 (1997).

51. А.М. Цирлин. Второй закон термодинамики и предельные возможности тепловых машин. ЖТФ, т. 69, в. 1, с. 140-142 (1999).

52. Г.Р. Иваницкий, А.Б. Медвинский, А.А. Деев, М.А. Цыганов. От "демона Максвелла" к самоорганизации процессов массопереноса в живых системах. УФН, т. 168, № 11, с. 1221 – 1233 (1998).

53. М.Б. Менский. Квантовая механика: новые эксперименты, новые приложения и новые формулировки старых вопросов. УФН, т. 171, № 4, с. 631-648 (2000).